\documentclass[a4paper, 12pt]{article}

\usepackage{tikz}
\usetikzlibrary{arrows, decorations.markings, cd, calc, shapes.geometric, decorations.pathmorphing}
\usepackage{latexsym, amsmath, amsfonts, amssymb, amsthm}
\usepackage{enumitem}
\usepackage[utf8]{inputenc}
\usepackage[american]{babel}
\usepackage{bbm}
\usepackage{caption}
\usepackage{cite}    
\usepackage[pdfencoding=auto,pagebackref]{hyperref}
\hypersetup{colorlinks, citecolor=[rgb]{.7,0,0}, linkcolor=[rgb]{0,0,0.7}, urlcolor=[rgb]{0,0,0.5}}
\usepackage{float}

\renewcommand{\baselinestretch}{1.2}
\numberwithin{equation}{section}

\newcommand{\be}{\begin{equation}} \newcommand{\ee}{\end{equation}}
\newcommand{\bea}{\begin{equation} \begin{aligned}} \newcommand{\eea}{\end{aligned} \end{equation}}
\newcommand{\ba}{\begin{array}} \newcommand{\ea}{\end{array}}

\newcommand*\pFqskip{8mu}
\catcode`,\active
\newcommand*\pFq{\begingroup
        \catcode`\,\active
        \def ,{\mskip\pFqskip\relax}%
        \dopFq
}
\catcode`\,12
\def\dopFq#1#2#3#4#5{%
        {}_{#1}F_{#2}\biggl[\genfrac..{0pt}{}{#3}{#4};#5\biggr]%
        \endgroup
}

\let\Im\relax \DeclareMathOperator{\Im}{Im}

\colorlet{algAc}{red!70!black}
\tikzset{algA/.style={algAc, thick}}
\colorlet{repMc}{violet!90!black}
\tikzset{repM/.style={repMc, thick}}
\tikzset{dashperp/.style={dash pattern = on 9 pt off 3 pt}}

\makeatletter
\def\blfootnote{\gdef\@thefnmark{}\@footnotetext}
\makeatother

\begin{document}

\thispagestyle{empty}
\begin{flushright}
\end{flushright}
\vspace{13mm}  
\begin{center}
{\huge Scalar QED in AdS}
\\[13mm]
{\large Ankur$^{1,2}$, Dean Carmi$^{3,4,5}$, Lorenzo Di Pietro$^{1,2}$,}

\bigskip
{\it

$^1$ INFN, Sezione di Trieste, Via Valerio 2, 34127 Trieste, Italy \\[.2em]
$^2$  Dipartimento di Fisica, Universit\`a di Trieste, Strada Costiera 11, 34151 Trieste, Italy\\[.2em]

$^3$ 
Department of Mathematics and Physics University of Haifa at Oranim, Kiryat Tivon
36006, Israel

$^4$  Department of Physics, Technion-Israel Inst. of Technology, Haifa 32000, Israel

$^5$  School of Physics and Astronomy, Tel Aviv University, Ramat Aviv 69978, Israel

}


\vspace{2cm}

{\parbox{16cm}{\hspace{5mm}
We consider scalar QED with $N_f$ flavors in AdS$_D$. For $D<4$ the theory is strongly-coupled in the IR. We use the spin 1 spectral representation to compute and efficiently resum the bubble diagram in AdS, in order to obtain the exact propagator of the photon at large $N_f$. We then apply this result to compute the boundary four-point function of the charged operators at leading order in $1/N_f$ and exactly in the coupling, both in the Coulomb and in the Higgs phase. In the first case a conserved current is exchanged in the four-point function, while in the second case the current is absent and there is a pattern of double-trace scaling dimension analogous to a resonance in flat space.  We also consider the BCFT data associated to the critical point with bulk conformal symmetry separating the two phases. Both in ordinary perturbation theory and at large $N_f$, in integer dimension $D= 3$ an IR divergence breaks the conformal symmetry on the boundary by inducing a boundary RG flow in a current-current operator.}}
\end{center}

\newpage
\pagenumbering{arabic}
\setcounter{page}{1}
\setcounter{footnote}{0}
\renewcommand{\thefootnote}{\arabic{footnote}}

{\renewcommand{\baselinestretch}{.88} \parskip=0pt
\setcounter{tocdepth}{2}
\tableofcontents}


\section{Introduction}
\label{sec: intro}

The geometry of (Euclidean) AdS$_D$ has several features that make it an ideal background to study quantum field theories (QFT) \cite{Callan:1989em}. It introduces a dimensionful parameter, the radius, which acts as an IR cutoff and can be used to probe the theory at different scales. Differently from other possible IR cutoffs, it preserves a large symmetry, namely the isometry group $SO(1,D)$. Moreover, it admits asymptotic observables, the correlators on the conformal boundary, on which the symmetry acts as the conformal group. These boundary correlators obey all the axioms of a $d=D-1$ dimensional conformal field theory (CFT), with the only exception of the existence of the stress-tensor operator. They are related to the S-matrix in the flat space limit \cite{Polchinski:1999ry, Giddings:1999jq, Gary:2009mi, Gary:2009ae, Fitzpatrick:2010zm, Okuda:2010ym, Penedones:2010ue, Fitzpatrick:2011jn, Fitzpatrick:2011hu, Raju:2012zr, Paulos:2016fap, Hijano:2019qmi, Komatsu:2020sag, Hijano:2020szl, Li:2021snj, Gadde:2022ghy, vanRees:2022zmr, Duary:2022pyv}. 

It is natural to ask how the bulk physics of the QFT maps to the boundary correlation functions. A good understanding of this dictionary paves the way to import the progress in CFT to massive quantum field theory, by applying the conformal bootstrap to the boundary correlators either with numerical \cite{Paulos:2016fap, Paulos:2016but, Paulos:2017fhb} or analytic techniques \cite{Mazac:2016qev, Mazac:2018mdx, Mazac:2018ycv, Paulos:2019gtx, Knop:2022viy}. This question is especially interesting in the regime of strong coupling. This motivated the study in \cite{Carmi:2018qzm} of strongly coupled theories of self-interacting scalars and fermions, namely the $O(N)$ and Gross-Neveu models, respectively, in the limit of large $N$ and finite coupling in a AdS$_D$ background. Other recent works on QFT in AdS studied how the bulk RG can be encoded in the boundary correlation functions \cite{Hogervorst:2021spa, Antunes:2021abs, Cordova:2022pbl, Levine:2023ywq, Meineri:2023mps}, studied thermal properties \cite{Kakkar:2022hub, Kakkar:2023gzu}, or considered the special case in which the bulk theory is conformal, as an efficient tool to study conformal defects \cite{Beccaria:2019dws, Beccaria:2019stp, Beccaria:2019mev, Beccaria:2019dju, Beccaria:2020qtk, Giombi:2021uae} or boundary conditions \cite{Herzog:2019bom, Giombi:2020rmc, Giombi:2021cnr}.

In this paper we apply the approach of \cite{Carmi:2018qzm} to a strongly coupled gauge theory. Asymptotically free gauge theories are a clear target to be studied using the AdS background \cite{Aharony:2012jf}, possibly via the bootstrap of the boundary correlators. It is therefore particularly important to understand how various gauge theory phenomena are encoded in the conformal correlators. We perform the first steps in this direction, studying the simple example of scalar QED (sQED) with $N_f$ flavors in the large $N_f$ limit. This theory is asymptotically free and has an interesting structure of phases for $2<D< 4$ ($D$ can be kept as a continuous parameter at large $N_f$). In flat space, it has a Coulomb phase and a Higgs phase separated by a second order phase transition, described by an interacting CFT. Both phases are gapless: in the Coulomb phase the massless excitation is the photon, while in the Higgs phase there are Goldstone bosons of the $\mathbb{CP}^{N_f-1}$ model. We place the theory in AdS$_D$ with Dirichlet-type boundary condition for the gauge field. These phases are still present in AdS$_D$ (and both are allowed for an intermediate range of $m^2$). In both phases we compute the four-point function of the charged operators created by the scalar electrons of the bulk theory, from which the dimensions of the exchanged operators can be extracted, for arbitrary values of the scaled gauge coupling $\alpha =e^2 N_f$. 

As an intermediate step, we compute the bubble diagram corresponding to the bulk two-point correlator of a conserved current in the free theory. To this end, we employ and further develop the technique of the spectral representation for two-point functions of a spinning operator \cite{Costa:2014kfa}.\footnote{While this paper was in preparation, \cite{Loparco:2023rug} appeared that contains the calculation of the analogous diagram in dS, with Wightman ordering, and also discusses in general the integral representation for spinning two-point functions that we use here. The calculation in dS is closely related to the one in AdS that we perform here: it amounts to replacing the AdS propagators with AdS harmonic functions. It should be possible to obtain the bubble we compute here from the dS one by performing two additional spectral integrals that convert the harmonic functions to AdS propagators. We thank Manuel Loparco for a discussion on this.}  The spectral representation allows us to readily resum the bubble diagrams and obtain the exact propagator of the photon at the leading order at large $N_f$. The four-point function is then expressed as an exchange diagram with this exact propagator. In the Coulomb phase, the spin 1 exchanged operators are: a conserved current with protected dimension, and the finite-coupling versions of the spin 1 double-trace operators of the matter fields, whose dimensions and OPE coefficient we can follow to finite values of $\alpha$ (there is a caveat for integer dimension $D=3$ that we discuss below). In the Higgs phase, the external operators are exactly marginal because the corresponding bulk fields are massless Goldstone bosons. A classical analysis in AdS would suggest that the current operator becomes non-conserved and gets an anomalous dimension. At finite coupling instead the only remnant of this non-conserved current is in a specific feature of the spectrum of the spin 1 double-trace operators, which is the AdS analogue of a resonance in flat space. Going to the deep IR with a tuned value of the mass-squared we reach a critical point with bulk conformal symmetry, corresponding to a BCFT in flat space via a Weyl rescaling, and we can extract the scaling dimensions of the spin 1 boundary operators appearing in the boundary OPE of the gauge field. 

Something special happens in the Coulomb phase in integer dimension $D=3$: the double-trace $j^\mu j_\mu$ of the boundary theory is classically marginal, and the corresponding coupling gets a non-trivial $\beta$ function triggered by the bulk gauge coupling. Therefore, the conformal symmetry of the boundary gets broken. This is true in ordinary perturbation theory \cite{Ren:2010ha, Faulkner:2012gt} and we explain that it remains true working at large $N_f$, finite coupling. We comment on the interpretation of this phenomenon as an IR divergence that is not cured by the AdS length scale, and offer a novel point of view on the computation of the $\beta$ function from the spectral representation of the propagator. This IR divergence persists when we tune the mass-squared of the scalar to the critical value, and hinders the possibility to define a conformal boundary condition for the IR CFT of 3d sQED by considering the RG in AdS with Dirichlet conditions for the gauge field. Therefore, when we talk about the boundary CFT in the Coulomb phase or at the bulk conformal point, we always refer to using a non-integer value of $D$ to regulate this divergence. 

The rest of the paper is organized as follows: in section \ref{sec:flat} we discuss scalar QED at large $N_f$ in flat space; this section is mostly a review, though typically only the critical point is discussed while here we also consider the observables away from criticality. In section \ref{sec:AdS} we compute the exact propagator of the gauge field in AdS, requiring the cancellation of spurious double-trace poles in the four-point function of the charged operators. Sections \ref{sec:CouAdS} and \ref{sec:HigAdS} contain an analysis of the spectrum in the Coulomb and Higgs phase, respectively, and also a discussion of the IR divergence in AdS$_3$ in the Coulomb phase. In section \ref{sec:CritAdS} we comment on the bulk critical point. The appendix discusses several properties of the spectral representation for the bulk two-point functions of conserved spin 1 operators.

\section{Phases of sQED in flat space}
\label{sec:flat}

 The Euclidean Lagrangian of sQED is
\begin{equation}\label{eq:Lag}
\mathcal{L} = \frac{1}{4 e^2} F_{\mu\nu} F^{\mu\nu}+m^{2} \phi^{a}\phi^{a*} +\left(D_{\mu} \phi^{a}\right)\left(D^{\mu} \phi^{a}\right)^{*}+\frac{\sigma}{\sqrt{N}}\left(\phi^{a} \phi^{a*}\right)-\frac{\sigma^{2}}{2 \lambda}
\end{equation}
where $\phi^{a}$ are $N_f$ complex scalar fields, $D_{\mu}= \partial_{\mu}+i A_{\mu}\,$, and $\sigma$ is a Hubbard-Stratonovich field with algebraic equation of motion $\sigma=\frac{\lambda}{\sqrt{N_f}} \phi^{a} \phi^{a*}$. Integrating out $\sigma$ gives a quartic interaction between the scalars. We will study the theory at large $N_f$ and for any value of the couplings $\lambda$ and $\alpha\equiv e^2 N_f$. Most of the formulas will be valid for any spacetime dimension $D=d+1$ but our focus will be in the range $2<D<4$ in which the theory is strongly coupled in the IR. Note that the continuous symmetries of this theory are a flavor symmetry $SU(N_f)$ rotating the scalar fields, and for integer $D$ a magnetic $U(1)$ $(D-3)$-form symmetry whose conserved current is $\frac{1}{2\pi}\star F$.

The first step in studying the dynamics of the theory is to minimize the effective potential and find the possible vacua/phases at large $N_f$ as a function of the parameters. At leading order at large $N_f$ the calculation of the effective potential as a function of the VEVs of $\sigma$ and $\phi^a$ is not affected by the presence of the gauge field and therefore the results are identical to those in the ungauged $O(2 N_f)$ model \cite{Coleman:1974jh}. In particular one finds that in the vacuum there is a VEV $\langle \sigma \rangle = \sqrt{N_f} \, \Sigma$ which has the effect of shifting the physical mass of the scalars to $M^2 = m^2 +\Sigma$. Both $m^2$ and $\Sigma$, which is itself a function of $m^2$, are UV divergent and scheme-dependent but their combination $M^2$ is not. The phase of the theory then depends on the value of $m^2$. For $m^2 > m^2_c$ we have a Coulomb phase with a massless gauge field that mediates a long range force $V(r) \sim e^2 \, r^{3-D}$ between massive charged particles with $M^2 > 0$. For $m^2 < m^2_c$ on the other hand we have a Higgs phase, the IR excitations are $2 N_f -2$ massless goldstone bosons parametrizing a $\mathbb{C P}^{N_f-1}$ sigma model, while the gauge-field becomes massive. Separating the two phases there is a second order phase transition at $m^2 = m^2_c$ described by a CFT.\footnote{We stress that these conclusions are valid in the regime of large $N_f$, though one can compute systematically $1/N_f$ corrections. The phases of this theory and of the tricritical theory at large $N_f$ were recently analyzed also in presence of a Chern-Simons term in \cite{DiPietro:2023zqn}. At small value of $N_f$ instead the transition might be first order, see e.g. section 5 of \cite{Gorbenko:2018ncu} for a discussion of the case $N_f = 2$, and also the relative size of $\lambda$ and $e^2$ can play an important role, e.g. in determining type I vs type II superconductivity in the Higgs phase for $N_f=1$, see e.g. \cite{Shifman:2022shi} for a textbook account. See \cite{Bonati:2020jlm, Bonati:2021vvs, Bonati:2021hzo, Bonati:2021wod, Bonati:2022oez, Bonati:2022srq, Bonati:2023vrf} for recent investigations of the phases of this theory at finite $N_f$ using lattice simulations.}

We will now consider each of these phases separately and compute some salient observables at leading order in the $1/N_f$ expansion.

\subsection{Coulomb phase}

In order to compute in $1/N_f$ perturbation theory we need the propagators of the fields $A_\mu$ and $\hat{\sigma} \equiv \sigma - \sqrt{N}_f \Sigma$ at order $1/N_f$, and exactly in the coupling constants. The calculation for $\hat{\sigma}$ at this order is not affected by the presence of the gauge field and therefore it takes the same form as for the $O(2 N_f)$ model \cite{Coleman:1974jh} which we will not repeat. 

\begin{figure}
\centering
\includegraphics[width=0.9\textwidth]{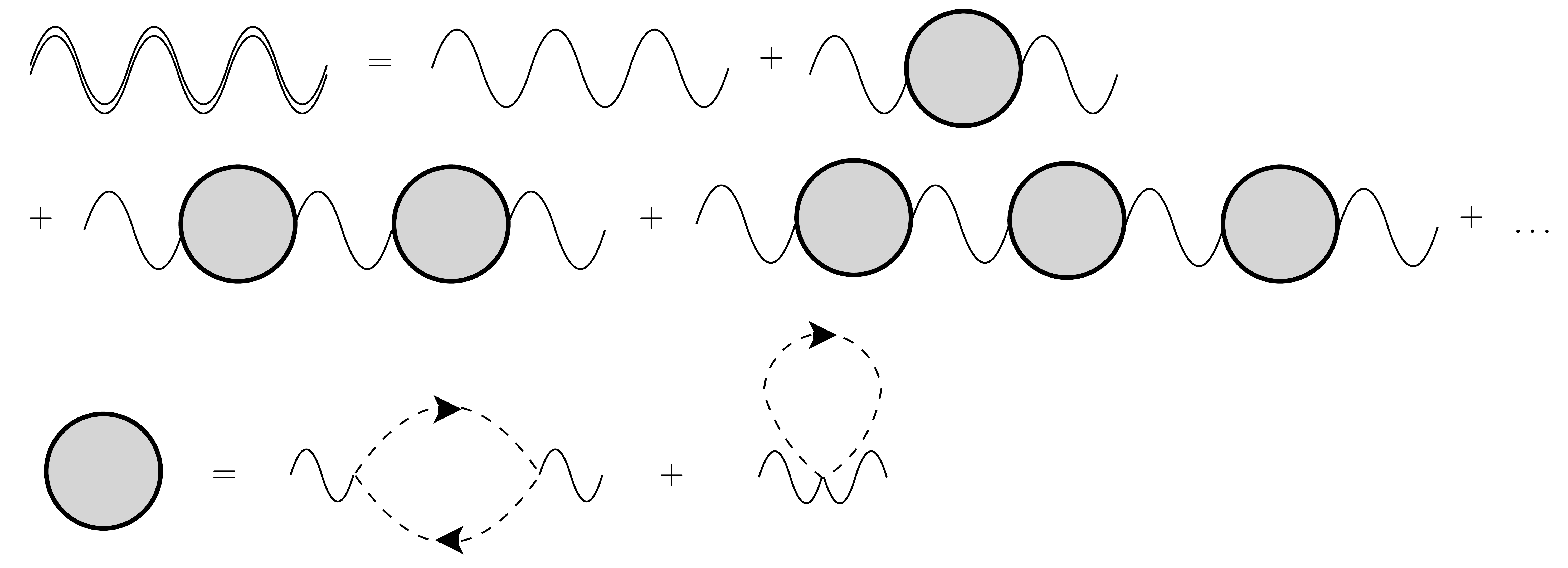}
\caption{The double wavy line is the propagator of the photon at leading order $\mathcal{O}(N_f^{-1})$ in the large $N_f$ expansion, and exactly in $\alpha = e^2 N_f$, while the single wavy line is the tree level propagator $\propto e^2$. The grey blob represents the 1-loop 1PI correction to the two-point function that is $\propto N_f$. The dashed line is the propagator of the complex scalars, with the arrow denoting the flow of charge.}\label{fig:bubblesum}
\end{figure}

For the gauge field, the exact propagator is given by the resummation of the 1PI bubble diagrams in figure \ref{fig:bubblesum}. 
Note that the 1PI bubble is the two-point function of the $U(1)$ conserved current in the theory of the $N_f$ free scalars, with the addition of the seagull contact term that ensures transversality, i.e. that $p^\mu \langle j_\mu(p) j_\nu(-p)\rangle = 0$. The result is
\begin{align}
\begin{split}\label{eq:spin1bubbleflat}
& \langle j_\mu(p) j_\nu(-p)\rangle  = - N_f B^{(1)}(p^2, M^2)\left(\delta_{\mu \nu} - \frac{p_{\mu} p_{\nu} }{p^{2}}\right)~,\\
\hspace{-0.15cm} B^{(1)}(p^2,M^2)  & = \frac{1}{D-1}\left[\left(p^{2}+4 M^{2}\right)B^{(0)}(p^2,M^2) - \frac{(4-2 D)}{(4 \pi)^\frac{D}{2}}\left(M^{2}\right)^{\frac{D}{2}-1} \Gamma\left(1-\frac{D}{2}\right)\right]~.
\end{split}
\end{align}
Here $B^{(0)}(p^2,M^2)$ denotes the spin 0 bubble function, i.e. the two point function of $\phi\phi^*$, given by
\begin{align}
\begin{split}
&B^{(0)}(p^2,M^2)  = \int \frac{d^D k}{(2 \pi)^{D}}\frac{1}{\left(k^{2}+M^{2}\right)((k+p)^{2}+M^{2})} \\
& = \frac{\Gamma \left(2-\frac{D}{2}\right) (M^2)^{\frac{D}{2}-2} }{(4 \pi )^{D/2} (3-D)}\left(\left(D-6-\frac{4 M^2}{p^2}\right) \, _2F_1\left(1,2-\frac{D}{2},\frac{1}{2},-\frac{p^2}{4 M^2}\right) \right. \\ &\hspace{4cm}\left. +\left(1+\frac{4 M^2}{p^2}\right) \, _2F_1\left(1,2-\frac{D}{2},-\frac{1}{2},-\frac{p^2}{4 M^2}\right)\right)~.
\end{split}
\end{align}

The tree-level propagator of the gauge field with gauge fixing Lagrangian 
\begin{equation}
\mathcal{L}_{g.f.} = \frac{1}{2 \xi} (\partial_\mu A^\mu)^2 ~,
\end{equation}
is
\begin{equation}
\langle A_\mu(p) A_\nu(-p) \rangle\vert_{\text{tree}} = \frac{e^2}{p^2}\left(\delta_{\mu \nu }-\frac{ p_\mu p_\nu}{p^{2}}\right)+\xi \frac{ p_{\mu} p_{\nu}}{p^4}~.
\end{equation}
The sum of bubble diagrams then becomes simply a geometric sum in the coefficient of the transverse projector, giving the following propagator at leading order at large $N_f$, and any fixed $\alpha=N_f e^2$ and $\zeta = N_f \xi$
\begin{equation}
\langle A_\mu(p) A_\nu(-p) \rangle\vert_{\frac{1}{N_f}} = \frac{1}{N_f}\left(\frac{\alpha}{p^2+ \alpha B^{(1)}(p^2,M^2)}\left(\delta_{\mu \nu }-\frac{ p_\mu p_\nu}{p^{2}}\right)+\zeta \frac{ p_{\mu} p_{\nu}}{p^4}\right)~.
\end{equation}
Note that the photon remains massless because $B^{(1)}(p^2,M^2) \underset{p\to 0}{\propto} p^2$.

\subsubsection{Scattering in the Coulomb phase}\label{eq:scatCou}

\begin{figure}
\centering
\includegraphics[width=0.8\textwidth]{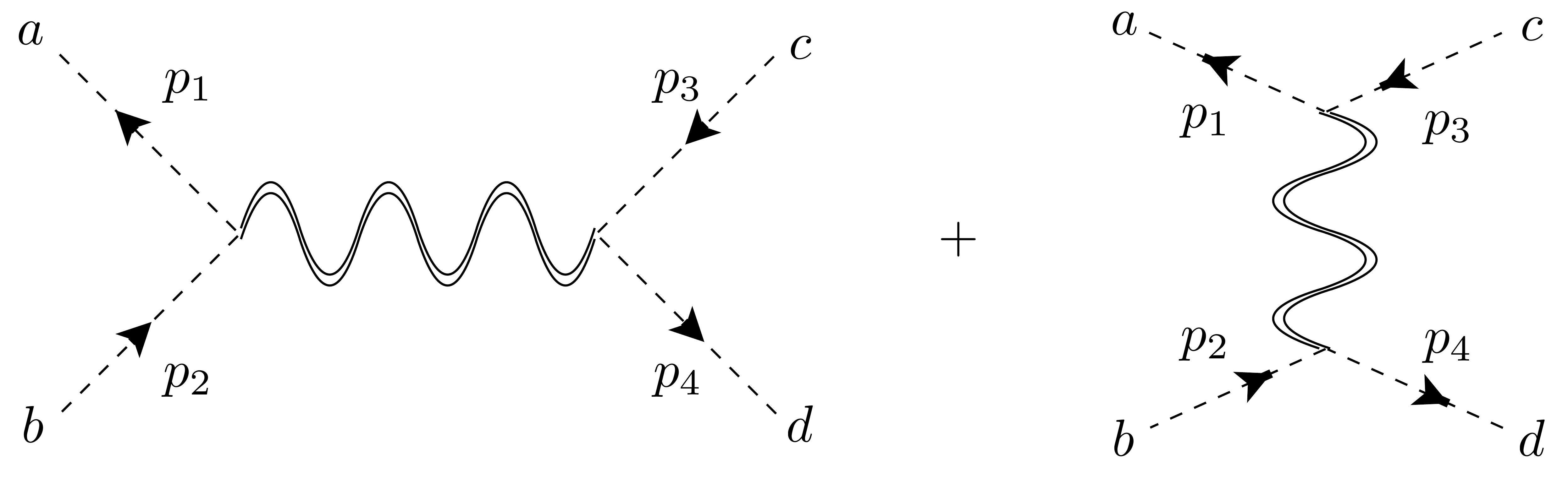}
\caption{Diagrams that compute the scattering amplitude at leading order $\mathcal{O}(N_f^{-1})$. The letters denote the $SU(N_f)$ flavor index and the momenta are all ingoing with $p_1 + p_2 +p_3+p_4 = 0$.}\label{fig:scatt}
\end{figure}

The simplest observable to compute is the scattering amplitude of the charged scalars $\phi^{*a} \phi^b \to \phi^{*c}\phi^d$. The contribution from the gauge field is given by the diagrams in figure \ref{fig:scatt}. 
It is immediate to write down the resulting amplitude using the exact photon propagator
\begin{align}
\begin{split}
& i \,T_{ab\to cd}  = i\,\frac{1}{N_f} \left(\delta^{ab}\delta^{cd} \,\mathcal{T}(s, t)+\delta^{ac}\delta^{bd}\, \mathcal{T}(t, s)  \right)~,\\
& i \mathcal{T}(s, t)   = -\frac{\alpha(s-4 M^2+2t)}{s - \alpha B^{(1)}(-s, M^2)}~.
\end{split}
\end{align}
The amplitude is crossing symmetric under simultaneous exchange of the flavor indices $b$ and $c$ and the Mandelstam variables $s$ and $t$. It also has analytic properties that are expected in the interacting theory, e.g. as a function of complex $s$ for fixed $t$ there is a pole at $s = 0$ due to the photon exchange and a two-particle branch-cut starting at $s = 4 M^2$. 

It is also interesting to check the unitarity of the amplitude, in particular after projecting to the singlet sector. Note that we can view the amplitude as the matrix element
\begin{equation}
i \,T_{ab\to cd} = \langle c, d | \, i \,T \, | a, b \rangle~,
\end{equation}
where the asymptotic two-particle states are normalized as
\begin{equation}
\langle a', b' | a, b \rangle = \delta^{a a'} \delta^{b b'} \times (\text{momentum conserving delta's})~.
\end{equation}
Therefore the unit normalized flavor singlet state is
\begin{equation}
 | \text{S} \rangle = \frac{1}{\sqrt{N}_f} \sum_a | a, a \rangle~,
\end{equation}
and the amplitude in the singlet sector is
\begin{align}
\begin{split}
& i \,T_{\text{S}\to \text{S}}  = i\,\mathcal{T}(s, t)+\mathcal{O}(N_f^{-1})~.
\end{split}
\end{align}
Note that $t$ only appears in the combination $s-4 M^2+2t = (s-4 M^2) \cos\theta$ in the numerator, where $\theta$ is the scattering angle. Therefore the decomposition in partial waves contains only spin $J=1$. In the normalization of \cite{Correia:2020xtr} we have
\begin{equation}
f^{\text{S}\to \text{S}}_{J=1}(s) = - \, \frac{\pi}{(16\pi)^{\frac{D-1}{2}}\Gamma(\frac{D+1}{2})}\frac{\alpha(s-4 M^2)}{s - \alpha B^{(1)}(-s,M^2)}~.
\end{equation}
Since this projection to the singlet sector is not suppressed by any small parameter, the full non-linear unitarity constraint applies to it, for any $\alpha$. In fact elastic unitarity is saturated,\footnote{Compared to \cite{Correia:2020xtr}, whose normalization we are using, there is an additional factor of 2 on the right-hand side because the external particle are complex scalars instead of identical real scalars.}
 i.e. for any $\alpha$ and any real $s>4M^2$ we have
\begin{equation}
2 \Im f^{\text{S}\to \text{S}}_1(s) = 2 \frac{(s-4 M^2)^{\frac{D-3}{2}}}{\sqrt{s}} |f^{\text{S}\to \text{S}}_1(s)|^2~.
\end{equation}
This can be easily checked using the following identity valid for real $s>4M^2$
\begin{align}
\begin{split}
\mathrm{Im}\,B^{(0)}(-s,M^2) = -\frac{ M^{D-4} }{2^{D+1} \pi ^{\frac{D-3}{2}} \Gamma \left(\frac{D-1}{2}\right)}\sqrt{\frac{4 M^2}{s}} \left(\frac{s}{4 M^2}-1\right)^{\frac{D-3}{2}}~.
\end{split}
\end{align}
The fact that at leading order in the $1/N_f$ expansion there is no particle production is a consequence of the fact that we are resumming only one-loop diagrams. Note that there is an additional contribution to the $\phi^{*a} \phi^b \to \phi^{*c}\phi^d$ amplitude from the exchange of the $\sigma$ field, i.e. from the scalar self-interaction, which however only contributes to the $J=0$ partial wave and similarly, when projected to the singlet sector, at leading order at large $N_f$ saturates elastic unitarity.

For $D\leq4$ at subleading order in the $1/N_f$ expansion we expect that IR divergences make the amplitude of the charged particles ill-defined in the Coulomb phase, see e.g. \cite{Galati:2021njb} for a recent discussion of IR divergences in sQED with $D=3$. Therefore one would need to consider some dressing of the asymptotic states, or replace the scattering amplitude with some inclusive observable.

\subsection{Higgs phase}

In this phase the minimization of the effective potential requires a non-zero VEV for $\phi^{a\,*}\phi^a = N_f \Phi^2$, and we have $M^2 = 0$, i.e. $\Sigma = - m^2$. Orienting the VEV of $\phi^a$ in the direction $a=N_f$ and labeling with $A=1,\dots, N_f-1$ the orthogonal directions, we expand the fields around the minimum as
\begin{align}
\begin{split}
&\phi^{A}=\pi^{A}~, \\
&\phi^{N}=\left(\sqrt{N_f} \Phi+\frac{\rho}{\sqrt{2}}\right) e^{i \frac{\theta}{\sqrt{2N_f} \Phi}}~,\\
&\sigma=- \sqrt{N_f} \, m^2 +\hat{\sigma}~.
\end{split}
\end{align}
The field fluctuations are: the Goldstone bosons $\pi^{A}$, the radial mode $\rho$ and the Hubbard-Stratonovich field $\hat{\sigma}$. Plugging in the Lagrangian \eqref{eq:Lag} we obtain
\begin{align}
\begin{split}\label{eq:LagHiggs}
\mathcal{L}   &=\frac{1}{4 e^2} F_{\mu\nu} F^{\mu\nu}+N_f \Phi^{2} A_{\mu} A^{\mu}+(D^\mu \pi^A) (D_\mu \pi^A)^*+\frac12\left(\partial_{\mu} \rho\right)^2+\frac{1}{2}\left(\partial_{\mu} \theta\right)^2\\
    &+\sqrt{N_f} \left(\frac{m^2}{\lambda}+\Phi^2 \right) \hat{\sigma} -\frac{\hat{\sigma}^{2}}{2 \lambda} + \sqrt{2} \,\Phi \, \hat{\sigma}\, \rho +\frac{\hat{\sigma}}{\sqrt{N_f}}\left(\frac{1}{2}\rho^2+ \pi^A\pi^{A\,*}\right)\\
    &+\sqrt{2 N_f} \,\Phi\, A_{\mu}\left(\partial^{\mu} \theta\right)+ \sqrt{2N_f}  \Phi \,\rho\, A_{\mu} A^{\mu}+2 \rho A_{\mu}\left(\partial^{\mu} \theta\right)+\frac{1}{2}\, \rho^{2} A_{\mu} A^{\mu}\\
    &+\frac{1}{\sqrt{2 N_f}\,\Phi} \rho^{2} A_{\mu}\left(\partial^{\mu} \theta\right)+\frac{1}{\sqrt{2 N_f}\,\Phi} \rho\left(\partial_{\mu} \theta\right)^2+\frac{1}{4 N_f \Phi^2}\,\rho^{2}\left(\partial_{\mu} \theta\right)^2~.
\end{split}
\end{align}

As expected we have a mass term for the gauge field. We can get rid of the mixing terms $A_\mu \partial^\mu \theta$ by introducing the gauge-fixing term 
\begin{equation}
\mathcal{L}_{g.f.} = \frac{N_f}{2 \zeta}\left(\partial_{\mu} A^{\mu} + \sqrt{\frac{2}{N_f}} \,\zeta\,\Phi\,\theta\right)^{2}~.
\end{equation}
With this choice, we can easily resum the 1PI diagrams in figure \ref{fig:bubblesum} to get the following photon propagator at leading order at large $N_f$
\begin{equation}
\langle A_\mu(p) A_\nu(-p) \rangle\vert_{\frac{1}{N_f}} = \frac{1}{N_f}\left(\frac{\alpha}{p^2+ m^2_A+\alpha B^{(1)}(p^2,0)}\left(\delta_{\mu \nu }-\frac{ p_\mu p_\nu}{p^{2}}\right)+\zeta \frac{ p_{\mu} p_{\nu}}{p^4}\right)~,
\end{equation}
where $m_A^2 = 2 e^2 \Phi^2$. Note that now only the $N_f-1$ massless fields $\pi^A$ run in the bubble loop. The bubble diagram at $M^2 = 0$ reads
\begin{equation}\label{eq:bubbleflatzeroMsq}
B^{(1)}(p^2,0) = -\frac{\pi  }{(16 \pi)^{\frac{D-1}{2}}\Gamma \left(\frac{D+1}{2}\right)\sin \left(\frac{\pi  D}{2}\right)} (p^2)^{\frac{D-2}{2}}~.
\end{equation}
Neglecting the bubble the massive photon is a stable particle corresponding to the pole at $p^2 = - m_A^2$. This value of $p^2$ on the negative real axis is precisely on the branch-cut of the power appearing in the bubble function. For $2<D< 4$ and any $\alpha$ the actual pole of the exact propagator is for complex values of $p^2$ and not in the first sheet.\footnote{The equation for the zero of the denominator is $p^2 + C(p^2)^\gamma = -m_A^2$, with $C>0$ and $0<\gamma < 1$. On the first sheet $\mathrm{Arg}(p^2) \in (-\pi,\pi)$ and $\mathrm{Arg}(C(p^2)^\gamma) = \gamma\, \mathrm{Arg}(p^2)$. Note that the solution cannot have $\mathrm{Arg}(p^2) = 0$ because in that case both $p^2$ and $C(p^2)^\gamma$ are positive real numbers. As a result, in the first sheet the arguments of $p^2$ and $C(p^2)^\gamma$ are either both in $(0,\pi)$ and their sum has a positive imaginary part, or both in $(-\pi,0)$ and their sum has negative imaginary part. In either case their sum cannot equal the real number $-m^2_A$. This proves that the solution is not in the first sheet.} The massive photon becomes a resonance, as expected given that it can decay to pions.

An observable in the Higgs phase is the scattering amplitude of the pions, that can be computed at leading order at large $N_f$ with the same techniques showed in section \ref{eq:scatCou}. Like we saw in the Coulomb phase, this amplitude is directly determined by the exact propagator of the photon, and therefore in this case it will contain a spin 1 resonance.  Like in the Coulomb phase, the $J=1$ partial amplitude in the singlet sector saturates elastic unitarity. An important difference with the Coulomb phase is that in the Higgs phase we do not expect any IR divergence, therefore this scattering amplitude remains an interesting observable of the theory also at subleading order in the $1/N_f$ expansion, or at finite $N_f$. 

\subsection{CFT at the phase transition}

For completeness let us now briefly review the evidence that at large $N_f$ there is a second order transition at $m^2 = m^2_c$, namely at $M^2 =0$ and $|\Phi|^2= 0$. The photon propagator with this value of the parameters and at leading order in the large $N_f$ expansion is 
\begin{equation}
\langle A_\mu(p) A_\nu(-p) \rangle\vert_{\frac{1}{N_f}} = \frac{1}{N_f}\left(\frac{\alpha}{p^2+\alpha B^{(1)}(p^2,0)}\left(\delta_{\mu \nu }-\frac{ p_\mu p_\nu}{p^{2}}\right)+\xi \frac{ p_{\mu} p_{\nu}}{p^4}\right)~,
\end{equation}
where $B^{(1)}(p^2,0)$ is the power of momentum in eq.~\eqref{eq:bubbleflatzeroMsq}. In the IR limit $(p^2)^{\frac{D-2}{2}} \ll \alpha$ the propagator approaches the $\alpha$-independent limit
\begin{equation}\label{eq:CFTprop}
\langle A_\mu(p) A_\nu(-p) \rangle\vert_{\frac{1}{N_f},\text{IR}} = \frac{1}{N_f}\left(\frac{1}{B^{(1)}(p^2,0)}\left(\delta_{\mu \nu }-\frac{ p_\mu p_\nu}{p^{2}}\right)+\xi \frac{ p_{\mu} p_{\nu}}{p^4}\right)~,
\end{equation}
from which we compute the two-point function of the gauge-invariant field strength operator
\begin{align}
\begin{split}
\langle \tfrac{1}{2\pi}F_{\rho\mu}(p) \tfrac{1}{2\pi} F_{\sigma\nu}(-p) \rangle\vert_{\frac{1}{N_f},\text{IR}}  & = \frac{C_F}{N_f}\frac{\delta_{\mu \nu }p_\rho p_\sigma - \delta_{\rho \nu }p_\mu p_\sigma -\delta_{\mu \sigma }p_\rho p_\nu +\delta_{\rho \sigma }p_\mu p_\nu }{(p^2)^{\frac{D-2}{2}}}~,\\
C_F & \equiv-\frac{(16 \pi)^{\frac{D-1}{2}}\Gamma \left(\frac{D+1}{2}\right)\sin \left(\frac{\pi  D}{2}\right)}{4\pi^3}~.
\end{split}
\end{align}
Note that $C_F > 0$ for $2<D<4$. In position space this correlator is
\begin{align}
\begin{split}
\langle \tfrac{1}{2\pi}F_{\rho\mu}(x) \tfrac{1}{2\pi} F_{\sigma\nu}(0) \rangle\vert_{\frac{1}{N_f},\text{IR}}   & = \frac{C_F}{N_f}\frac{16}{(4\pi)^{\frac{D}{2}}\Gamma(\tfrac{D}{2}-1)}\frac{I_{\mu\nu}I_{\rho\sigma}- I_{\rho\nu} I_{\mu\sigma}}{(x^2)^2}~,\\
I_{\mu\nu} & \equiv \delta_{\mu\nu}-2\frac{x_\mu x_\nu}{x^2}~.
\end{split}
\end{align}
This takes precisely the form of the correlator for a two-form primary operator of scaling dimension $\Delta_F = 2$ in a $D$ dimensional CFT. Note that the unitarity bound for a two-form is $\Delta \geq \mathrm{max}(2,D-2)$, so in the range $2<D<4$ this bound is saturated by $F$, reflecting the existence of the null operator corresponding to the Bianchi identity $d F = 0$. In integer $D$ the hodge dual $\star F$ gives the conserved current for a $(D-3)$-form symmetry, but only in $D=3$ this conserved current is compatible with the unitarity of the CFT.\footnote{For $D=4$ both $\Delta =2$ and $\Delta = D-2$ are saturated, so unitarity would require both $d F = 0$ and $d\star F = 0$, but the second identity does not hold due to the coupling to matter fields.}

At leading order at large $N_f$, and restricting to local operators, the CFT is the product of two decoupled sectors, the mean field theory of the field strength operator, and the free CFT of the matter fields, restricted to the singlet sector of the $U(1)$ gauge symmetry and with the conserved current removed $J^\mu = 0$. Corrections to the CFT data can be computed systematically in $1/N_f$ expansion, e.g. using a diagrammatic approach with the exact propagator \eqref{eq:CFTprop} and the standard interaction vertices involving the matter fields. In addition in $D=3$ there are also local monopole operators, for which the $1/N_f$ expansion is less straightforward but has also been developed, see e.g. \cite{Pufu:2013eda, Dyer:2015zha, Chester:2022wur}.

\section{sQED in AdS}
\label{sec:AdS}

We will now study large $N_f$ scalar QED in AdS space. The theory is defined by the same Lagrangian as in eq. \eqref{eq:Lag}. The possible relevant curvature coupling $R \,\phi^a \phi^{a\,*}$ is absorbed in the definition of the coupling $m^2$. It will be convenient to keep the spacetime dimension as an arbitrary parameter, so will work in AdS$_{D=d+1}$.

\subsection{Exact photon propagator}
Just like in flat space, the first step towards the large $N_f$ solution is to obtain the exact large $N_f$ propagator of the gauge field. For this purpose, it is convenient to adopt the spectral representation\footnote{In this paper we use the term ``spectral representation'' to refer to the integral representation which expands a bulk two-point function in the continuous basis of eigenvalues of the AdS Laplacian, i.e. harmonic functions. Recently the same term \cite{Meineri:2023mps} was used to refer to boundary OPE expansion of a bulk two-point function. The two representations are related by closing the contour in the integral representation and rewriting the integral as a discrete sum over poles. While the representation in \cite{Meineri:2023mps} is indeed a closer analogue to the K\"all\'en-Lehman spectral representation in flat space, we keep using the term to refer to the integral representation to be consistent with \cite{Carmi:2018qzm}, and for lack of a better term.} of the propagator \cite{Costa:2014kfa}. Imposing a Dirichlet boundary condition for the gauge field at the conformal boundary of AdS, i.e. the standard boundary condition that gives rise to a global symmetry on the boundary, the propagator in ordinary perturbation theory has the following spectral representation
\begin{align}\label{eq:propA}
\begin{split}
 \langle A_M(X) A_N(Y) \rangle_{\text{pert. theory}}  & \equiv G^{(1)}_{MN}(X,Y)\\
& = \int_{-\infty}^{+\infty} d\nu \,\frac{e^2}{\nu^2 + (\frac{d}{2}-1)^2} \,\Omega^{(1)}_{\nu\,MN}(X,Y) + \nabla^X_M\nabla^Y_N L(u)~.
\end{split}
\end{align}
Here we are using the embedding coordinates $X,Y \in \mathbb{R}^{1,d+1}$ with mostly plus signature convention and $X^2 = Y^2 = -1$. Therefore the indices $N,M$ run from $0$ to $d+1$ and they are constrained to be transverse in order to lie on the tangent space, i.e. $X^M A_M(X) = 0$ and similarly in the point $Y$, see \cite{Costa:2014kfa} for more details.

The function $\Omega^{(1)}_{\nu\,MN}(X,Y)$ is a spin 1 harmonic function in AdS, i.e. it is an eigenfunction of the Laplacian on vectors
\begin{align}
\square^X \Omega^{(1)}_{\nu\,MN}(X,Y) = \left(\nu^2+\frac{d^2}{4} +1\right) \Omega^{(1)}_{\nu\,MN}(X,Y)~,
\end{align}
with no singularity at coincident points. It is symmetric under exchange $X\leftrightarrow Y$ together with $M\leftrightarrow N$. Moreover, it is a transverse function
\begin{align}
\nabla^X_M \Omega^{(1)\,M}_{\nu~~~\,N}(X,Y) = 0~.
\end{align}
Therefore we see that the propagator \eqref{eq:propA} has a transverse part proportional to the harmonic function, and a longitudinal gauge-dependent part given by the derivative of an arbitrary function $L$ of the distance between the two points, parametrized by $u =-1 - X\cdot Y$. The familiar family of $R_\xi$ gauges, parametrized by a real parameter $\xi$, corresponds to the choice
\begin{equation}
L(u) = e^2\, \xi \, G^{(0)}_d \star G^{(0)}_d (u) = e^2\, \xi  \int_{-\infty}^{+\infty} d\nu\,\frac{1}{(\nu^2 +\frac{d^2}{4})^2} \, \Omega^{(0)}_\nu (u)~.
\end{equation}
Here $G^{(0)}_\Delta$ denotes the AdS$_{d+1}$ propagator of a scalar field whose corresponding boundary operator has scaling dimension $\Delta$, and choosing $\Delta = d$ corresponds to a massless scalar field, the symbol $\star$ denotes the convolution, and $\Omega^{(0)}_\nu$ is the scalar harmonic function.

The transverse harmonic function $\Omega^{(1)}_{\nu\,MN}(X,Y)$ can be used more generally to write down a spectral representation for an arbitrary bulk two-point function of a spin one conserved operator. The details of this transform, and how to obtain the Fourier transform in the flat-space limit, are explained in the appendix \ref{app:SpecRep}. In particular, for the two-point function of the $U(1)$ conserved current of a single complex scalar we have
\begin{equation}
\langle J_M(X) J_N(Y) \rangle = - \int_{-\infty}^{+\infty} d\nu \,B^{(1)}(\nu) \,\Omega^{(1)}_{\nu\,MN}(X,Y)~.
\end{equation}
$B^{(1)}(\nu)$ is a function of $\nu$ that we will have to determine. For $N_f$ complex scalars there is an additional overall factor of $N_f$. This two-point function is precisely the one-loop 1PI correction to the propagator of the photon. The contribution of the seagull interaction fixes the contact term to ensure that the two-point function is transverse even at separated points, as $\Omega^{(1)}_{\nu\,MN}$ is. 

Just like in flat space, the exact propagator at leading order at large $N_f$ is obtained by summing diagrams with an arbitrary number of insertions of this 1PI one-loop correction proportional to $N_f$. Thanks to the properties of $\Omega^{(1)}_{\nu\,MN}$ under convolution \cite{Costa:2014kfa}, this sum becomes geometric in the $\nu$ variable. Note also that the convolution of $\Omega^{(1)}_{\nu\,MN}$ with the longitudinal part of the propagator vanishes, because we can integrate by parts and use that $\Omega^{(1)}_{\nu\,MN}$ is transverse.\footnote{Since the diagrams are given by integrals over the whole AdS space, we can only drop the longitudinal contribution thanks to the fact that  $\Omega^{(1)}_{\nu\,MN}$ is transverse even at coincident points.} As a result we get
\begin{align}\label{eq:propAN}
\begin{split}
 & \langle A_M(X) A_N(Y) \rangle_{\text{large $N_f$}}  \\
& = \frac{1}{N_f}\int_{-\infty}^{+\infty} d\nu \,\frac{\alpha}{\nu^2 + (\frac{d}{2}-1)^2 + \alpha B^{(1)}(\nu)} \,\Omega^{(1)}_{\nu\,MN}(X,Y) + \nabla^X_M\nabla^Y_N L(u)~,
\end{split}
\end{align}
where $\alpha = e^2 N_f$ is kept fixed in the limit.

\subsection{Boundary four-point function}

We will now use \eqref{eq:propAN} to compute the boundary four-point function of the charged operators dual to the complex scalar. Denoting the boundary operators with the same letter $\phi^a$ that we use for the bulk fields, the four-point function at leading order is just the trivial disconnected contribution
\begin{align}
\begin{split}
& \langle \phi^a(P_1) \phi^{*\,b}(P_2) \phi^{*\,c}(P_3) \phi^{d}(P_4) \rangle \vert_{O(N_f^0)} \\
& = \delta^{ab} \delta^{cd} \frac{1}{(-2 P_1\cdot P_2)^{\Delta}(-2 P_3\cdot P_4)^{\Delta}} + \delta^{ac} \delta^{bd} \frac{1}{(-2 P_1\cdot P_3)^{\Delta}(-2 P_2\cdot P_4)^{\Delta}}~,
\end{split}
\end{align}
where $m^2 = \Delta(\Delta -d)$. Since the scalar is complex, there is no u-channel contribution, and as a result the OPE decomposition contains double-trace operators of any integer spin $J$, even and odd, with scaling dimensions $2 \Delta + 2n + J$, with $n$ non-negative integers. Note that as long as $\Delta > \frac{d}{2}-1$ there are no spin 1 conserved operators. This might seem at odds with the fact that there is a global symmetry under which $\phi^a$ are charged, but actually the boundary theory is not local and therefore Noether's theorem does not apply.

Next, we consider the first connected contribution at order $O(N_f^{-1})$. This is the exchange diagram of the photon, for which we use the exact propagator \eqref{eq:propAN}. The result is
\begin{align}
\begin{split}
& \langle \phi^a(P_1) \phi^{*\,b}(P_2) \phi^{*\,c}(P_3) \phi^{d}(P_4) \rangle \vert_{O(N_f^{-1})}  = \delta^{ab} \delta^{cd} g_{12\vert 34}+ \delta^{ac} \delta^{bd}g_{13\vert 24} ~,
\end{split}
\end{align}
where
\begin{align}
\begin{split}
& g_{ij\vert kl}   =\frac{1}{N_f}  \int_{-\infty}^{+\infty} d\nu \,\frac{\alpha}{\nu^2 + (\frac{d}{2}-1)^2 + \alpha B^{(1)}(\nu)} \\ & 4 \int_{X,Y} \, K_\Delta(P_i,X) \,i\nabla^X_M K_\Delta(P_j,X)  \, K_\Delta(P_k,Y) (-i) \nabla^Y_N K_\Delta(P_l,Y) \,\Omega^{(1)\,MN}_\nu(X,Y)~.
\end{split}
\end{align}
We are using the shorthand notation $\int_{X,Y} = \int d^{d+1} X\int d^{d+1} Y$ for the integral over the bulk points. $K_\Delta(P,X)$ is the bulk-to-boundary propagator of the scalar field
\begin{align}
\begin{split}
K_\Delta(P,X)  =\frac{ \sqrt{\mathcal{C}_\Delta}}{(-2 X\cdot P)^\Delta}~,\\
\mathcal{C}_\Delta  \equiv \frac{\Gamma(\Delta)}{2\pi^{\frac{d}{2}} \Gamma(\Delta -\frac{d}{2}+1)}~,
\end{split}
\end{align}
where the normalization comes from taking a canonical normalization for the bulk field, and normalizing to 1 the two-point function of the boundary operator. Note that thanks to gauge-invariance only the transverse part of the propagator contributes to the four-point function.

Next we use that \cite{Costa:2014kfa}
\begin{align}
\begin{split}
& \int_{X,Y}\, K_\Delta(P_i,X) \,i\nabla^X_M K_\Delta(P_j,X)  \, K_\Delta(P_k,Y) (-i) \nabla^Y_N K_\Delta(P_l,Y) \,\Omega^{(1)\,MN}_\nu(X,Y) \\
& \hspace{-1cm}=\frac{1}{(-2 P_i\cdot P_j)^\Delta(-2 P_k\cdot P_l)^\Delta} \frac{1}{8 \pi^\frac{d}{2} \Gamma(\Delta)^2 \Gamma(1-\frac{d}{2}+\Delta)^2} \mathcal{F}^{(1)}_{\frac{d}{2}+i\nu}(u,v)\\
& \hspace{-1cm}= \frac{1}{(-2 P_i\cdot P_j)^\Delta(-2 P_k\cdot P_l)^\Delta} \frac{1}{8 \pi^\frac{d}{2} \Gamma(\Delta)^2 \Gamma(1-\frac{d}{2}+\Delta)^2}  \left(C_\nu \mathcal{K}^{(1)}_{\frac{d}{2}+i\nu}(u,v)  + (\nu \to -\nu) \right)~,
\end{split}
\end{align}
where
\begin{equation}
C_\nu \equiv  \frac{\Gamma \left(\frac{d}{4}+\frac{i \nu }{2}+\frac{1}{2}\right)^4 \Gamma \left(-\frac{d}{4}+\Delta \pm \frac{i \nu }{2}+\frac{1}{2}\right)^2 }{ 2\pi \left(\frac{d}{2}+i \nu -1\right)\Gamma (i \nu ) \Gamma \left(\frac{d}{2}+i \nu +1\right)}~.
\end{equation}
Here $\mathcal{F}^{(1)}_{\frac{d}{2}+i\nu}$ is the spin 1 conformal partial wave, and $\mathcal{K}^{(1)}_{\frac{d}{2}+i\nu}$ is the spin 1 conformal block, which are functions of the scaling dimension $\frac{d}{2}+i\nu$ of the exchanged operator, and of the conformally invariant cross-ratios $u$ and $v$.\footnote{We use the same convention of \cite{Costa:2012cb} for the conformal partial wave and the conformal blocks, with the following map between symbols: $F_{\text{there}} =\mathcal{F}_{\text{here}}$ and $G_{\text{there}} =\mathcal{K}_{\text{here}}$.}

As a result the four-point function is
\begin{align}
\begin{split}
& g_{ij\vert kl}   =\frac{1}{N_f}  \frac{1}{(-2 P_i\cdot P_j)^\Delta(-2 P_k\cdot P_l)^\Delta} \frac{1}{2 \pi^{\frac{d}{2}} \Gamma(\Delta)^2 \Gamma(1-\frac{d}{2}+\Delta)^2}  \\
&\int_{-\infty}^{+\infty} d\nu \,\frac{\alpha}{\nu^2 + (\frac{d}{2}-1)^2 + \alpha B^{(1)}(\nu)} \left(C_\nu \mathcal{K}^{(1)}_{\frac{d}{2}+i\nu}(u,v)  + (\nu \to -\nu) \right)~.
\end{split}
\end{align}
Closing the contour of the $\nu$ integral in the lower-half plane for the $\mathcal{K}^{(1)}_{\frac{d}{2}+i\nu}$ term, and in the upper-half plane for $\mathcal{K}^{(1)}_{\frac{d}{2}-i\nu}$, we pick the contributions from the poles of the coefficient functions that generate the OPE expansion of the correlator. Note that at the tree level in usual perturbation theory $\alpha \ll 1$ we can neglect the bubble contribution and we have simply an exchange diagram of the photon. The corresponding pole at $\nu = \pm i (\frac{d}{2}-1)$ correspond to the exchange of a conserved current operator in the OPE. We see that the boundary current, that was absent in the theory of charged scalars with a bulk global symmetry, appears when we gauge the symmetry in the bulk with Dirichlet boundary conditions for the gauge field. The conserved current is still present in the OPE even at finite $\alpha$ as long as the bubble function in the denominator does not shift the position of the corresponding pole, namely
\begin{equation}\label{eq:gicond}
B^{(1)}\left(\pm i \left(\tfrac{d}{2}-1\right)\right) = 0~.
\end{equation}
This can be seen as a condition of unbroken gauge invariance on the bubble function. We will later use it to fully determine the function $B^{(1)}\left(\nu\right)$.

\subsection{Bootstrapping the spin 1 bubble}
We will now compute the bubble function $B^{(1)}\left(\nu\right)$ not by a direct calculation of the diagram, but rather by imposing a certain self-consistency condition on the four-point function. The method follows closely the one used in \cite{Carmi:2018qzm} to compute the spin 0 bubble function.

Consider the projection of the four-point function to the s-channel singlet sector of the flavor symmetry $SU(N_f)$. This projection simply amounts to contracting with $\frac{1}{N_f^2}\delta^{ab} \delta^{cd}$, giving
\begin{align}
\begin{split}
 \frac{1}{N_f^2}\langle \phi^a(P_1) \phi^{*\,a}(P_2) \phi^{*\,b}(P_3) \phi^{b}(P_4) \rangle & = \frac{1}{(-2 P_1\cdot P_2)^{\Delta}(-2 P_3\cdot P_4)^{\Delta}} \\
&\hspace{-2cm} + \frac{1}{N_f}\left( \frac{1}{(-2 P_1\cdot P_3)^{\Delta}(-2 P_2\cdot P_4)^{\Delta}} + g_{12\vert 34}\right) +\mathcal{O}(N_f^{-2})~.
\end{split}
\end{align}
The t-channel disconnected contribution and the s-channel connected one enter at the same order $\frac{1}{N_f}$. The factor $\Gamma \left(-\frac{d}{4}+\Delta \pm \frac{i \nu }{2}+\frac{1}{2}\right)^2$ in the connected diagram $g_{12\vert 34}$ has a double pole at the positions $\nu =\nu^{\pm}_n\equiv \pm i\left( 2\Delta + 2n +1-\frac{d}{2}\right)$, with $n$ a non-negative integer. In ordinary perturbation theory $\alpha \ll 1$, the term involving the function $B^{(1)}\left(\nu\right)$ in the denominator can be neglected, and these double poles have the effect of producing an $\mathcal{O}(\alpha)$ anomalous dimension for the spin 1 double-trace operators of dimension $2\Delta +2n+1$. 

On the other hand, at order $1/N_f$ and finite $\alpha$, the exchange of the exact propagator of the gauge field produces anomalous dimensions for these operators that are non-trivial functions of $\alpha$. These can only arise from zeroes of the denominator $\nu^2 + (\frac{d}{2}-1)^2 + \alpha B^{(1)}(\nu)$. At the same time, the exchange of these operators from the disconnected contribution needs to be canceled by a contribution of opposite sign from $g_{12\vert 34}$. This can happen only if $B^{(1)}(\nu)$ has a single pole precisely at the location of the double-pole in the numerator, so that the full $\nu$ integrand of $g_{12\vert 34}$ has only a single pole at that location. This gives the condition
\begin{align}
\begin{split}\label{eq:cancpoles}
& c^2_{n,J=1} \\ & = 2\pi i\, \mathrm{Res}\left.\left[2 \frac{1}{2 \pi^{\frac{d}{2}} \Gamma(\Delta)^2 \Gamma(1-\frac{d}{2}+\Delta)^2} \frac{\alpha}{\nu^2 + (\frac{d}{2}-1)^2 + \alpha B^{(1)}(\nu)} C_\nu\right]\right\vert_{\nu = \nu_n^-}~,
\end{split}
\end{align}
where the factor of 2 inside the parenthesis comes from the equal contributions of the two terms $\mathcal{K}^{(1)}_{\frac{d}{2}\pm i\nu}$, and $c_{n,J}$ are the OPE coefficients for a complex scalar GFF, which in our normalization of the conformal blocks read
\begin{equation}
c^2_{n,J} = \frac{2^J (\Delta -\tfrac{d}{2}+1)_n^2 (\Delta)^2_{n+J}}{J! n! (J+\tfrac{d}{2})_n(2\Delta +n-d+1)_n (2\Delta +2n +J-1)_J (2\Delta +n +J-\tfrac{d}{2})_n}~.
\end{equation}
Denoting the behavior of the bubble near the pole as
\begin{equation}\label{eq:poleb}
B^{(1)}(\nu) \underset{\nu \sim \nu_n^-}{\sim} \frac{b^{(1)}_n}{i(\nu -\nu_n^-)} + ...~,
\end{equation} 
we can solve \eqref{eq:cancpoles} to determine
\begin{align}
\begin{split}
& b_n^{(1)} \\ & \hspace{-0.15cm}= \frac{ \Gamma \left(\frac{d}{2}+n+1\right) \Gamma (n+\Delta +1) \Gamma \left(-\frac{d}{2}+n+\Delta +\frac{1}{2}\right) \Gamma \left(-\frac{d}{2}+n+2 \Delta +1\right)}{ (4\pi)^{\frac{d}{2}} \Gamma \left(\frac{d}{2}+1\right) \Gamma (n+1) \Gamma \left(n+\Delta +\frac{3}{2}\right) \Gamma \left(-\frac{d}{2}+n+\Delta +1\right) \Gamma (-d+n+2 \Delta +1)}~.
\end{split}
\end{align}

In ordinary perturbation theory, i.e. in an expansion in $\alpha$, the bubble function appears in the numerator as a loop correction to the photon exchange diagram. At any finite order in perturbation theory we cannot have new operators appearing in the spectrum, but rather we can only generate a series of corrections to the OPE data of the GFF theory. As a result, the singularities in \eqref{eq:poleb} are the only singularities of $B^{(1)}(\nu)$ in the complex plane. 

If in addition the function $B^{(1)}(\nu)$ would decay at infinity in the complex plane, by a simple contour argument the function would be uniquely fixed in terms of the location of the poles and the residues. However in generic dimension $B^{(1)}(\nu)$ does not decay, and this manifests in a divergence of the sum over poles with the prescribed residues. This is due to (bulk) UV divergences in the loop that computes the bubble. The summand (symmetrized under $\nu\to-\nu$) behaves as
\begin{equation}
B^{(1)}_n(\nu) \equiv \frac{2 i \nu_n^{-} \,b^{(1)}_n}{\nu^2 -(\nu_n^-)^2
} \underset{n\to \infty}{\sim} ~ -\frac{n^{d-2}}{(4\pi)^\frac{d}{2} \Gamma(\frac{d}{2}+1)}(1+\dots)~,
\end{equation}
where the dots denote a series of $1/n$ corrections, such that the coefficients of the $1/n^{2k}$ and $1/n^{2k+1}$ corrections are (even) polynomials in $\nu$ of degree $\nu^{2k}$. In any $d$ we can make the sum convergent by subtracting sufficient terms, say $2m$, in the Taylor expansion of the summand $B^{(1)}_n(\nu)$ around $\nu = 0$, which contains only even powers. After resumming the resulting convergent series, we account for the subtraction by adding a polynomial in $\nu$ of degree $2m$ with arbitrary coefficients.\footnote{The same procedure can be derived from the fact that $B^{(1)}(\nu)$ behaves at infinity as
\begin{equation}
B^{(1)}(\nu) \underset{|\nu| \to \infty}{\propto} |\nu|^{d-1}~.
\end{equation}
This growth implies that the contour argument determines $B^{(1)}(\nu)$ when $d<1$, while for $1\leq d <3$ we can only use it to determine $B^{(1)}(\nu) - B^{(1)}(0)$, for $3\leq d < 5$ we can only use it to determine $B^{(1)}(\nu)- B^{(1)}(0)-\frac{1}{2}\nu^2 B^{(1)''}(0)$, and so on. The behavior for $\nu \to \infty$ corresponds to the flat space limit in momentum space with $\nu \sim L p$, see the appendix \ref{app:SpecRep}, and therefore can be fixed by computing the behavior of the spin 1 bubble in $\mathbb{R}^{d+1}$. However this limit also requires to scale $\Delta\sim L m \to \infty$. Since we cannot prove that the leading power in the growth at large $\nu$ does not depend on $\Delta$ (though a posteriori this will turn out to be true) we prefer to use the argument based on the structure of the series in the main text.} 

From the structure described above we see that no subtraction is needed only for $d<1$. In the more interesting range $1\leq d < 3$ we have
\begin{equation}
B^{(1)}(\nu)\vert_{1\leq d < 3} = \sum_{n=0}^\infty \left[B^{(1)}_n(\nu)-B^{(1)}_n(0)\right] + a_0~,
\end{equation}
where the infinite sum is now convergent, but $a_0$ is a constant that remains undetermined.  We have the further constraint \eqref{eq:gicond} coming from the condition of gauge invariance, and we can use it to fix $a_0$. We get
\begin{equation}
B^{(1)}(\nu)\vert_{1\leq d < 3} = \sum_{n=0}^\infty \left[B^{(1)}_n(\nu)-B^{(1)}_n(i\left(\tfrac{d}{2}-1\right))\right]~.
\end{equation}
Note that, if one computes the loop with a choice of regulator, the coefficient $a_0$ is UV divergent if the regulator does not preserve gauge invariance (e.g. a sharp cutoff). The coupling that reabsorbs the UV divergence in $a_0$ is in fact the mass of the gauge field.  In the range $3\leq d < 5$ we need to perform one more subtraction to get a convergent sum
\begin{equation}
B^{(1)}(\nu)\vert_{3\leq d < 5} = \sum_{n=0}^\infty \left[B^{(1)}_n(\nu)-B^{(1)}_n(0)-\frac{\nu^2}{2} B^{(1)''}_n(0)\right] + a_0 + a_1(\nu^2 + (\tfrac{d}{2}-1)^2)~.
\end{equation}
and there are two undetermined constants $a_{0,1}$. We can again impose the gauge-invariance condition \eqref{eq:gicond} to fix $a_0$, obtaining
\begin{align}
\begin{split}
B^{(1)}(\nu)\vert_{3\leq d < 5} & = \sum_{n=0}^\infty \left[B^{(1)}_n(\nu)-B^{(1)}_n(i\left(\tfrac{d}{2}-1\right))-\frac{\nu^2 + (\tfrac{d}{2}-1)^2}{2}B^{(1)''}_n(0)\right] \\ & + a_1(\nu^2 + (\tfrac{d}{2}-1)^2)~.
\end{split}
\end{align}
However in this case the coefficient $a_1$ remains undetermined, and in fact computing explicitly the loop with a UV regulator one would find that $a_1$ is UV divergent, even with a gauge-invariant regulator. The UV divergence in $a_1$ is reabsorbed in the gauge coupling, which indeed is finite in the range $1\leq d < 3$ but needs to be renormalized in the range $3\leq d < 5$. More subtractions are needed if $d$ is further increased.

The final sums simplify when $d$ is even. For $d=2$ (i.e. AdS$_3$) we obtain
\begin{equation}
B^{(1)}(\nu)\vert_{d=2} = \frac{\nu  \left[-2 (2 \Delta -3 ) \nu +\left(\nu^2+ 4 (\Delta -1)^2\right) \left(i \,\psi\left(\Delta -\frac{i \nu }{2}\right)-i \,\psi\left(\Delta +\frac{i \nu }{2}\right)\right)\right]}{16 \pi  \left(\nu ^2+1\right)}~,
\end{equation}
where $\psi(x)$ denotes the digamma function. For $d=4$ (i.e. AdS$_5$) we obtain
\begin{align}
\begin{split}
& B^{(1)}(\nu)\vert_{d=4} = \frac{\nu ^2+1}{2048 \pi ^2 \nu  \left(\nu ^2+4\right)} \left[ (\Delta -1) (2 \Delta -5) (2 \Delta -7) \nu  \left(4-3 \nu ^2\right) \right. \\
& \left. -4 \left(\nu^2 +(2 \Delta -3)^2\right) \left(\nu^2 +(2 \Delta -5)^2\right) \left(i \,\psi \left(\Delta -\tfrac{i \nu }{2}-\tfrac{1}{2}\right)-i \psi \left(\Delta +\tfrac{i \nu }{2}-\tfrac{1}{2}\right)\right)\right] \\
& + \tilde{a}_1 (\nu ^2+1)  ~,
\end{split}
\end{align}
where the tilde denotes that we reabsorbed a $\Delta$-dependent constant in the undetermined coefficient. As a check of the result, we can compute the flat-space limit by taking $\nu = L p$, $p$ being the modulus of the momentum in flat space, and $\Delta = L m$, and sending $L\to \infty$, see the appendix \ref{app:SpecRep}. We find that the AdS results approach the flat space answer for the bubble computed with a dimreg regulator in eq. \eqref{eq:spin1bubbleflat}, up to a polynomial in the momentum when $d\geq 3$ which reflects the ambiguity in the choice of the regulator (which we left unspecified in AdS).

We can also write the sum in generic $d$ in terms of a generalized hypergeometric function. In the rest of the paper we focus on 
the range $1\leq d < 3$ in which case the expression reads
\begin{align}\label{eq:Bd}
& B^{(1)}(\nu)\vert_{1\leq d < 3} = \mathcal{B}(\nu) - \mathcal{B}(i(\tfrac{d}{2}-1))~,\nonumber\\
&\mathcal{B}(\nu)\equiv \frac{\pi ^{\frac{1-d}{2}} 2^{-2 \Delta } \nu ^2 \Gamma (\Delta +1) \Gamma \left(2 \Delta -\frac{d}{2}+1\right) \, }{\left(\Delta -\frac{d}{4}+\frac{1}{2}\right) \Gamma \left(\Delta +\frac{3}{2}\right) \Gamma \left(\Delta-\frac{d}{2} +1\right)^2\left(\nu ^2+4 \left(\Delta -\frac{d}{4}+\frac{1}{2}\right)^2\right) } \\ 
&\pFq{7}{6}{\frac{d}{2}+1,\Delta +1,\Delta -\frac{d}{2}+\frac{1}{2},2 \Delta -\frac{d}{2}+1,\Delta -\frac{d}{4}+\frac{1}{2},\Delta -\frac{d}{4}+\frac{i \nu }{2}+\frac{1}{2},\Delta -\frac{d}{4}-\frac{i \nu }{2}+\frac{1}{2}}{\Delta +\frac{3}{2},\Delta -\frac{d}{2}+1,2 \Delta -d+1,\Delta -\frac{d}{4}+\frac{3}{2},\Delta -\frac{d}{4}+\frac{i \nu }{2}+\frac{3}{2},\Delta -\frac{d}{4}-\frac{i \nu }{2}+\frac{3}{2}}{1}~.\nonumber
\end{align}
Having fixed the form of the exact propagator of the photon, we now study the physical observables that we can extract from it in the various phases of the theory.

\section{Coulomb Phase in AdS}
\label{sec:CouAdS}

For $m^2 \geq m^2_{c,1}$ there is stable minimum of the AdS effective potential at $\phi^a = 0$ and with a non-zero expectation value $\propto \Sigma$ for the Hubbard-Stratonovich field, which gives a physical mass-squared $M^2 = m^2 + 2 \Sigma$ to the scalar fluctuations above the Breitenlohner-Freedman (BF) bound \cite{Breitenlohner:1982jf}. The analysis of the effective potential at leading order at large $N_f$ is not affected by the gauge field and therefore we will not repeat it here but simply refer to the case of the self-interacting scalars \cite{Carmi:2018qzm}, from which one can also read off the (scheme-dependent) value of $m^2_{c,1}$. This phase of the gauge theory is the Coulomb phase, in which the photon mediates a long-range force between the scalars. We concentrate in the range $1\leq d <3$ in which the theory is strongly-coupled at large distances. We assume Dirichlet boundary conditions for the gauge field. 

The observables we will consider are the scaling dimensions of the spin 1 boundary operators that are exchanged in the connected four-point of the charged operators, at the leading order in the $1/N_f$ expansion. Equivalently, these are the operators that appear in the boundary channel expansion of the bulk two-point function of the gauge field. Setting the AdS scale $L=1$, they depend on two parameters, the gauge coupling $\alpha$ and the mass-squared $M^2$, which we will trade with the scaling dimension $\Delta$ of the boundary charged operator.

\subsection{$1\leq d < 3\,,~d\neq 2$: scaling dimensions from weak to strong coupling}

\begin{figure}
\centering
\begin{minipage}{.45\textwidth}
\centering
\includegraphics[width=\textwidth]{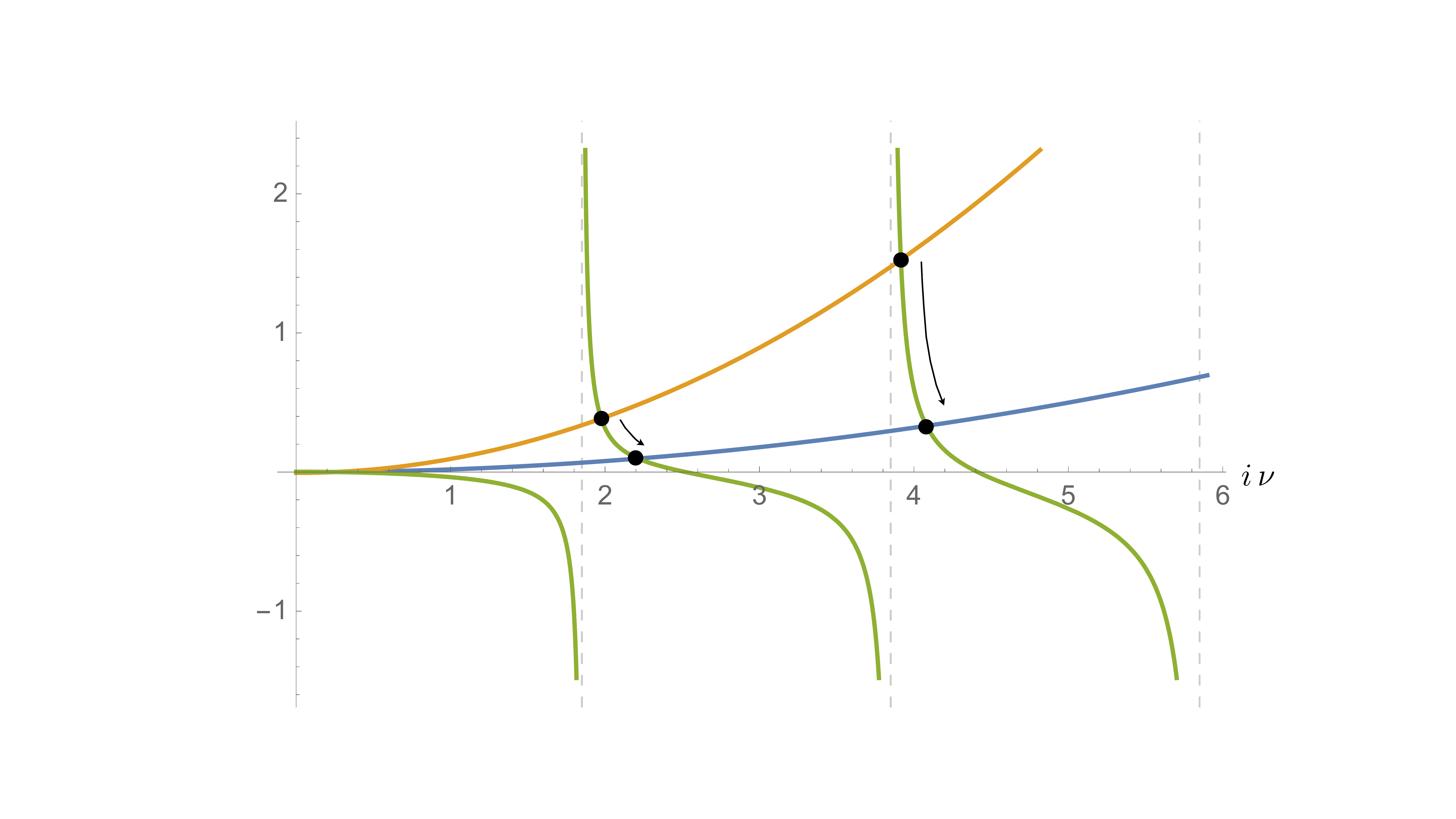}
\end{minipage}
\hspace{1cm}\begin{minipage}{.45\textwidth}
\centering
\includegraphics[width=\textwidth]{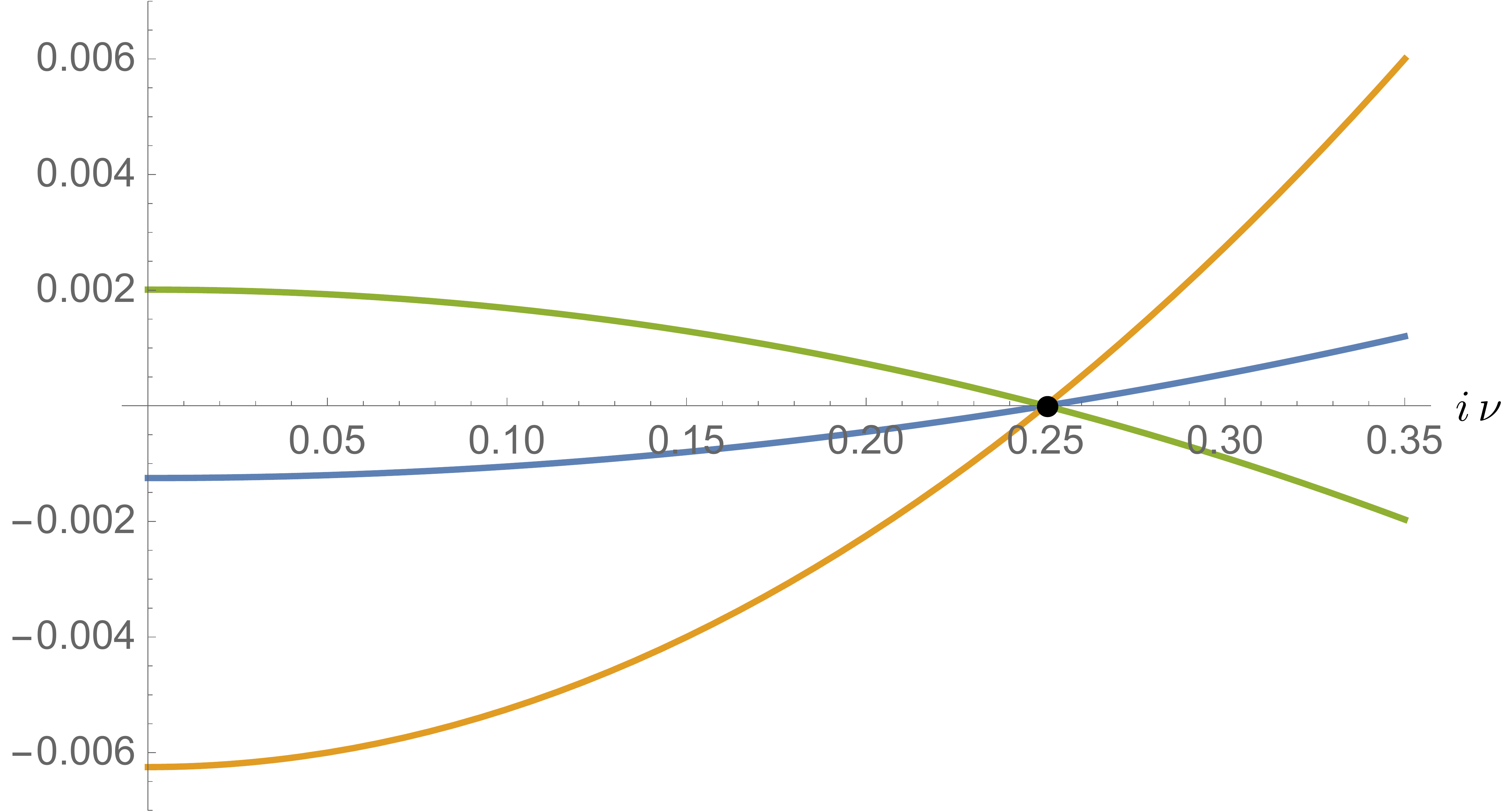}
\end{minipage}
\caption{On the left: Plot of $B^{(1)}(\nu)$ (green line) on the negative imaginary $\nu$ axis, together with $-\frac{1}{\alpha}(\nu^2 + (\tfrac{d}{2}-1)^2)$ for $\alpha = 10$ (orange curve) and $\alpha = 50$ (blue curve) in units of the AdS radius. We have taken $d= 5/2$ and $\Delta = 21/20$ for the external operators. The intersections are highlighted with black dots, the corresponding values of $\nu$ give the scaling dimensions of spin 1 operators via $\Delta = \frac{d}{2}+i\nu$. The arrows denote the direction of increasing coupling constant $\alpha$. The dashed vertical lines correspond to the spin 1 double-trace operators in the free theory. On the right: Zoom of the previous plot near the origin. There all the curves $-\frac{1}{\alpha}\left(\nu^2 + \left(\frac{d}{2}-1\right)^2 \right)$ for any $\alpha$ intersect $B^{(1)}(\nu)$ at the point $i\nu = \frac{d}{2}-1$. The corresponding operator is the conserved current of the global symmetry.}\label{fig:zeroesd}
\end{figure}

We first consider $d\neq 2$ in order to regulate the IR divergence that appears in AdS$_3$. The spectrum of the spin 1 boundary operators is determined by the poles of the exact propagator \eqref{eq:propAN} in the complex $\nu$ plane, i.e. by the zeroes of the denominator
\begin{equation}
\frac{1}{\alpha}\left(\nu^2 + \left(\frac{d}{2}-1\right)^2 \right) + B^{(1)}(\nu) = 0~.
\end{equation}
The solutions $\{\nu^*_n\}_{n\geq 0}$ are located on the negative imaginary $\nu$ axis, and they correspond to the exchange of an operator with scaling dimension $\Delta_n = \frac{d}{2}+i\nu^*_n$ (since the function is symmetric under $\nu\to-\nu$ we could equivalently look at the positive imaginary axis). We cannot find a closed form expression for these solutions, but we can easily visualize them by plotting the function $B^{(1)}(\nu)$ in eq. \eqref{eq:Bd} as a function of $\nu$ along the negative imaginary axis, for a given $\Delta$ and $d$. This is showed in figure \ref{fig:zeroesd}. We see that for small values of $\alpha$ the scaling dimensions approach their minimal values $\Delta^{(0)}_n = 2\Delta + 2n +1$ which are just the values in the free theory. The anomalous dimension increases monotonically as a function of the coupling, with no level crossing. The maximum value of $\Delta_n$ is still separated by a gap from $\Delta^{(0)}_{n+1}$ and is reached in the limit $\alpha \to \infty$, corresponding to the zeroes of $B^{(1)}(\nu)$. In addition to this tower of solutions, for any value of $\alpha$ there is also an additional zero at $i\nu^* = \tfrac{d}{2}-1$ which corresponds to the conserved current, as follows from the condition of gauge invariance \eqref{eq:gicond} that we used to fix the bubble function.

\begin{figure}
\centering
\begin{minipage}{.4\textwidth}
\centering
\includegraphics[width=\textwidth]{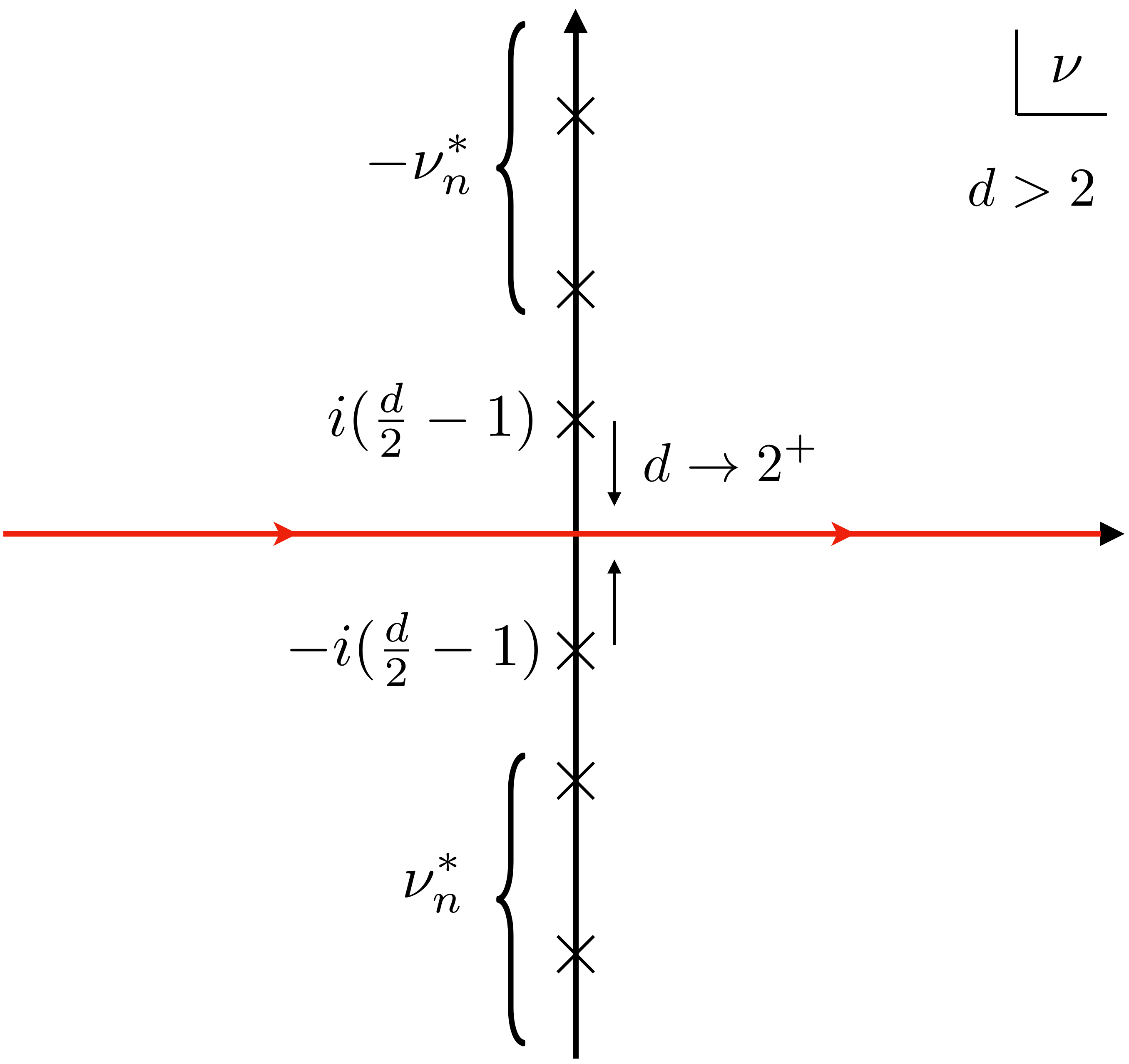}
\end{minipage}
\hspace{2cm}\begin{minipage}{.4\textwidth}
\centering
\includegraphics[width=\textwidth]{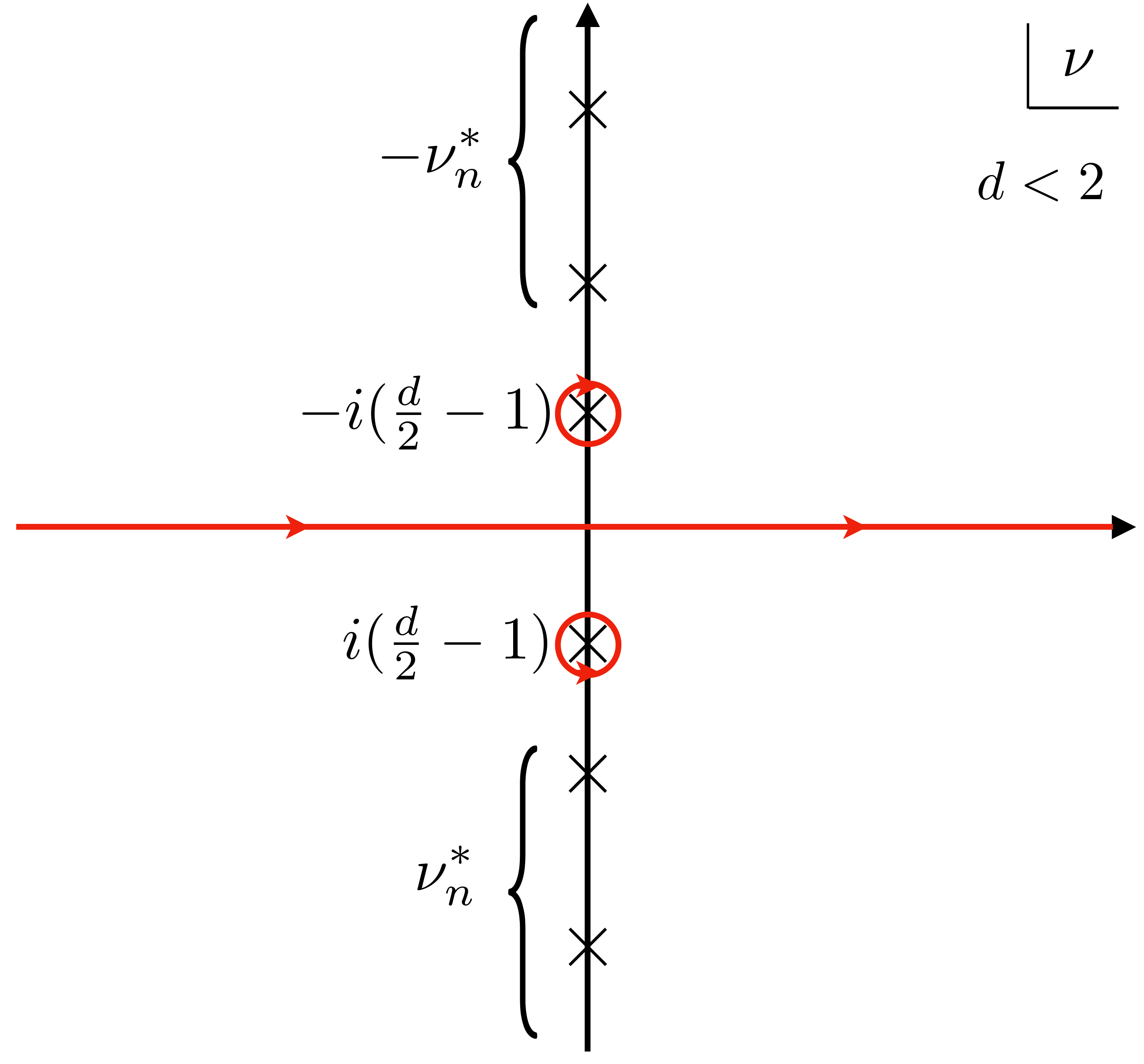}
\end{minipage}
\caption{Poles (crosses) and integration contour (red curve) of the spectral representation of the photon propagator in the complex $\nu$ plane. The poles at $\pm\nu_n^*$ gives the finite coupling version of the double-trace operators arising from the matter fields. The poles at $\pm i (\tfrac{d}{2}-1)$ give the conserved current. For $d<2$ they cross the contour and we need to add to the contour two circles surrounding them.}\label{fig:polecross}
\end{figure}

Note that as we go from $d>2$ to $d<2$ the conserved current poles in the upper- and lower-half $\nu$ plain cross the integration contour on the real $\nu$ axis, as illustrated in figure \ref{fig:polecross}. As a result to ensure continuity in $d$ the contour needs to be changed by adding circles surrounding these two poles, similarly to what is done for scalar AdS propagators with alternate boundary conditions. This is related to the fact that the two boundary modes of the vector field in AdS
\begin{equation}\label{eq:moded}
A_\mu \underset{z\to 0}{\sim}  z^{d-2} \,e^2\, j_\mu +  a_\mu+ \dots~,
\end{equation}
exchange dominance as we go from $d>2$ to $d<2$. Here we are using Poincar\'e coordinates $(z, x^\mu)$, $\mu=1,\dots,d$, with boundary at $z=0$,  $j_\mu$ denotes the boundary conserved current and $a_\mu$ the boundary gauge field. The dots denote subleading contributions from descendants, and also from higher dimensional operators when the gauge field is coupled to matter. The Dirichlet boundary condition sets $a_\mu = 0$.

The contribution from the piece of the contour surrounding the pole naively requires evaluating $\Omega^{(1)}_{\nu\,MN}$ at $\nu=i(\frac{d}{2}-1)$, however the harmonic function itself is singular there, it has a single pole. One should then evaluate the residue at the resulting double pole, but alternatively we observe that the residue of $\Omega^{(1)}_{\nu\,MN}$ is longitudinal, namely
\begin{equation}\label{eq:longdiv}
\Omega^{(1)}_{\nu\,MN}(X,Y) \underset{\nu\to i(\frac{d}{2}-1)}{\sim} \frac{\nabla^X_M\nabla^Y_N F(u)}{\nu -i(\frac{d}{2}-1) } + \bar{\Omega}^{(1)}_{MN}(X,Y)~. 
\end{equation}
Recalling that the $\nu$ integral computes the two-point function of the gauge field, thanks to gauge-invariance we can ignore the longitudinal piece and simply consider the finite term denoted as $\bar{\Omega}^{(1)}_{MN}(X,Y)$.

\subsection{$d=2$: IR divergence and breaking of conformal invariance}\label{eq:IRdiv}

In the limit $d\to2$ the poles associated to the conserved current pinch the contour, see figure \ref{fig:polecross}, making the propagator of the photon with Dirichlet boundary condition singular in AdS$_3$. This singularity arises in the spectral representation of the photon propagator from the behaviour of the $\nu$ integral in eq. \eqref{eq:propA} around $\nu = 0$
\begin{equation}
\sim \int \frac{d\nu}{\nu^2}~.
\end{equation}
The analogy between the $\nu$ integral and momentum space integrals in flat space suggests the interpretation of this divergence as an IR divergence in the bulk of AdS. On the other hand from the point of view of the boundary conformal theory this manifests like a UV divergence and can be reabsorbed in a running coupling, which leads to a breaking of conformal invariance. 

\begin{figure}
\centering
\includegraphics[width=0.35\textwidth]{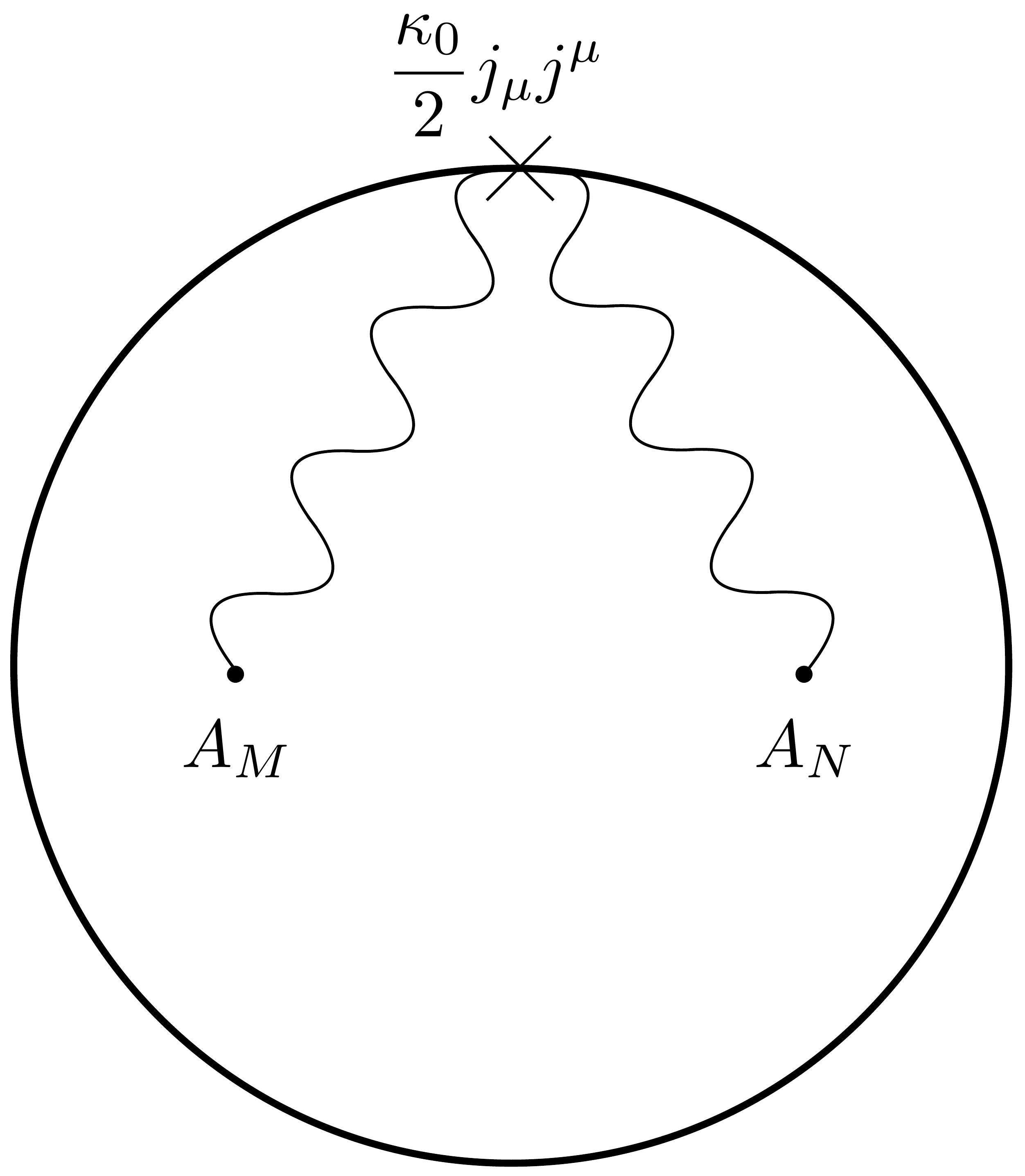}
\caption{The correction to the propagator at leading order in the boundary current-current interaction. The wavy lines are bulk-to-boundary propagators, and the boundary point is integrated over.}\label{fig:CorrPropPert}
\end{figure}

We can understand the relation between this divergence and the running of the coupling as follows: adding the marginal interaction on the boundary
\begin{equation}
\delta S_\text{boundary} =\frac{\kappa_0}{2} \, \int d^d x \,\hat{j}_\mu \hat{j}^\mu~,
\end{equation}
gives an additional contribution to the propagator of the gauge field, represented by the diagram in fig. \ref{fig:CorrPropPert}.  Using $d\neq 2$ as a regulator, the contribution of this diagram is expected to be proportional to the value of the harmonic function at the pole that is pinching the contour, namely $\bar{\Omega}^{(1)}_{MN}(X,Y)$ up to longitudinal terms, and therefore the pole $\frac{1}{d-2}$ can be absorbed by a renormalization of the coupling $\kappa_0 \propto \frac{\mu^{d-2}}{d-2}$. This in turn gives rise to a $\beta$ function for $\kappa$.

Instead of computing the diagram, a simple way to obtain this $\beta$ function is by looking at the boundary condition of the gauge field \cite{Ren:2010ha, Faulkner:2012gt} (similar results in the scalar case were discussed earlier in \cite{Klebanov:1999tb, Witten:2001ua}). In $d=2$ the boundary conserved current appears as the coefficient of a logarithmic mode in the near boundary expansion of the gauge field
\begin{equation}\label{eq:mode2}
A_\mu \vert_{d=2} \underset{z\to 0}{\sim}  \log z\,e^2\,\hat{j}_\mu +  a_\mu+ \dots~.
\end{equation}
Using dimreg and comparing \eqref{eq:mode2} with \eqref{eq:moded} we see that the $d=2$ current $\hat{j}_\mu$ is related to $d$ dimensional counterpart as
\begin{equation}\label{eq:rescj}
j_\mu = \frac{1}{d-2} \,\hat{j}_\mu + \mathcal{O}(1)~.
\end{equation}
The resulting pole in the near-boundary expansion must be reabsorbed by a re-definition of the constant mode
\begin{equation}\label{eq:kapparen}
a_\mu = -\kappa_0 \,\hat{j}_\mu~,~~ \kappa_0 = \left(\frac{e^2}{d-2} + \kappa(\mu)\right)\mu^{2-d}~.
\end{equation}
This mixed boundary condition corresponds to turning on the double-trace coupling $\frac{\kappa_0}{2} \, \hat{j}_\mu \hat{j}^\mu$\cite{Witten:2001ua} which is classically marginal in $d=2$.\footnote{To fix the normalization of the current-current coupling, one needs to consider the $d$-dimensional boundary action in the presence of the source. With the normalization in \eqref{eq:moded} one finds that the coupling between the source and the current is $-\int d^d x \, (d-2) \, a_\mu j^\mu$ which in the limit $d\to 2$ gives $-\int d^d x \, a_\mu \hat{j}^\mu$.} Here $\kappa_0$ denotes the bare dimreg coupling, $\kappa(\mu)$ the renormalized coupling, and $\mu$ is the dimreg scale. We then obtain the leading order $\beta$ function for the coupling
\begin{equation}
0 = \frac{d \kappa_0}{d \log \mu} \Rightarrow  \beta_\kappa = \frac{d \kappa}{d\log\mu} = -e^2 +\mathcal{O}(\kappa^2, e^4, e^2 \kappa) ~,
\end{equation}
which depends only on the bulk coupling and not on $\kappa$ itself at leading order.

We can use the same logic to compute the $\beta$ function of $\kappa$ at leading order for small $\kappa$ and large $N_f$, at any value of $\alpha$. The only difference is that the boundary OPE coefficient $e^2$ of $\hat{j}_\mu$ in the expansion of the gauge field gets now replaced by the coefficient of $\frac{1}{\nu^2}$ in the expansion around $\nu\to 0$ of the full propagator \eqref{eq:propAN}. In this way we get
\begin{equation}
\beta_\kappa\vert_{\text{large $N_f$}} = - \frac{1}{N_f} \frac{\alpha }{1+\frac{\alpha}{8 \pi }  \left((3-2 \Delta )+2 (\Delta -1)^2 \psi(\Delta )\right)}   +\mathcal{O}(\kappa^2, N_f^{-2}, N_f^{-1} \kappa)~.
\end{equation}

As a result, both in perturbation theory and at large $N_f$ the Dirichlet boundary condition in AdS$_3$ in the Coulomb phase does not preserve the isometry and it does not allow to define a set of boundary conformal correlator. It would be interesting to explore the existence of fixed points for the coupling $\kappa$ with some appropriate scaling of the coupling with $e^2$ or $\frac{1}{N_f}$. As the derivation did not use any detail of the matter sector, the existence of this boundary running couplings is a generic phenomenon for 3d gauge theories in AdS, and persists even for the pure gauge theory. An important exception is the case with a Chern-Simons term in the Lagrangian.

\section{Higgs Phase in AdS}
\label{sec:HigAdS}

For $m^2 \leq m^2_{c,2}$ the AdS effective potential has minimum with $\phi^a = \sqrt{N_f} \Phi^a \neq 0$ and with vanishing mass-squared of the scalar fluctuations $M^2 = 0$. This is the Higgs phase, in which the gauge field gets a mass $m^2_A = 2 e^2 \Phi^2$, and the $N_f -1$ massless scalar fluctuations correspond to Goldstone bosons for the spontaneous breaking of the flavor symmetry. We refer again to \cite{Carmi:2018qzm} for the discussion of the effective potential, we simply recall that $m^2_{c,1} <  m^2_{c,2}$ and as a result in AdS there is a range of parameters in which the Coulomb and the Higgs phase are both possible. 

The Lagrangian for the Higgs phase in AdS is the same as the one in flat space in eq. \eqref{eq:LagHiggs}.  The corresponding large $N_f$ propagator of the gauge field is
\begin{align}\label{eq:propHiggs}
\begin{split}
 & \langle A_M(X) A_N(Y) \rangle_{\text{large $N_f$, Higgs phase}}  \\
& = \frac{1}{N_f}\int_{-\infty}^{+\infty} d\nu \,\frac{\alpha}{\nu^2 + (\Delta_A -\frac{d}{2})^2 + \alpha B^{(1)}(\nu)\vert{_{\Delta = d}}} \,\Omega^{(1)}_{\nu\,MN}(X,Y) + \nabla^X_M\nabla^Y_N L(u)~,
\end{split}
\end{align}
where $m^2_A =(\Delta_A -1)(\Delta_A -d+1)$. Goldstone bosons in AdS are associated to the existence of a conformal manifold of boundary theories, on which the bulk global symmetry acts \cite{Carmi:2018qzm}. In this Higgs phase, a $U(1)$ factor of the spontaneously broken symmetry is gauged, and in the boundary conformal theory this means that the marginal operators are charged under the would-be $U(1)$ symmetry, which consequently is explicitly broken by the marginal couplings. The current operator is therefore not protected anymore, and classically it would get a scaling dimension $\Delta_A$ above the unitarity bound $\geq d-1$.

The observables we will consider are the scaling dimensions of the spin 1 boundary operators that are exchanged in the connected four-point of the Goldstone bosons $\pi^A$, at the leading order in the $1/N_f$ expansion. Equivalently, these are the operators that appear in the boundary channel expansion of the bulk two-point function of the massive gauge field. Setting the AdS scale $L=1$, they depend on two parameters, the gauge coupling $\alpha$ and the mass-squared $m^2_A$, which we will trade with the scaling dimension $\Delta_A$. Unlike the Coulomb phase, having generated a mass for the gauge field there is no IR divergence in this phase, and there is no need to discuss $d=2$ separately. 

\subsection{Spin 1 resonance in AdS}

\begin{figure}
\centering
\includegraphics[width=0.6\textwidth]{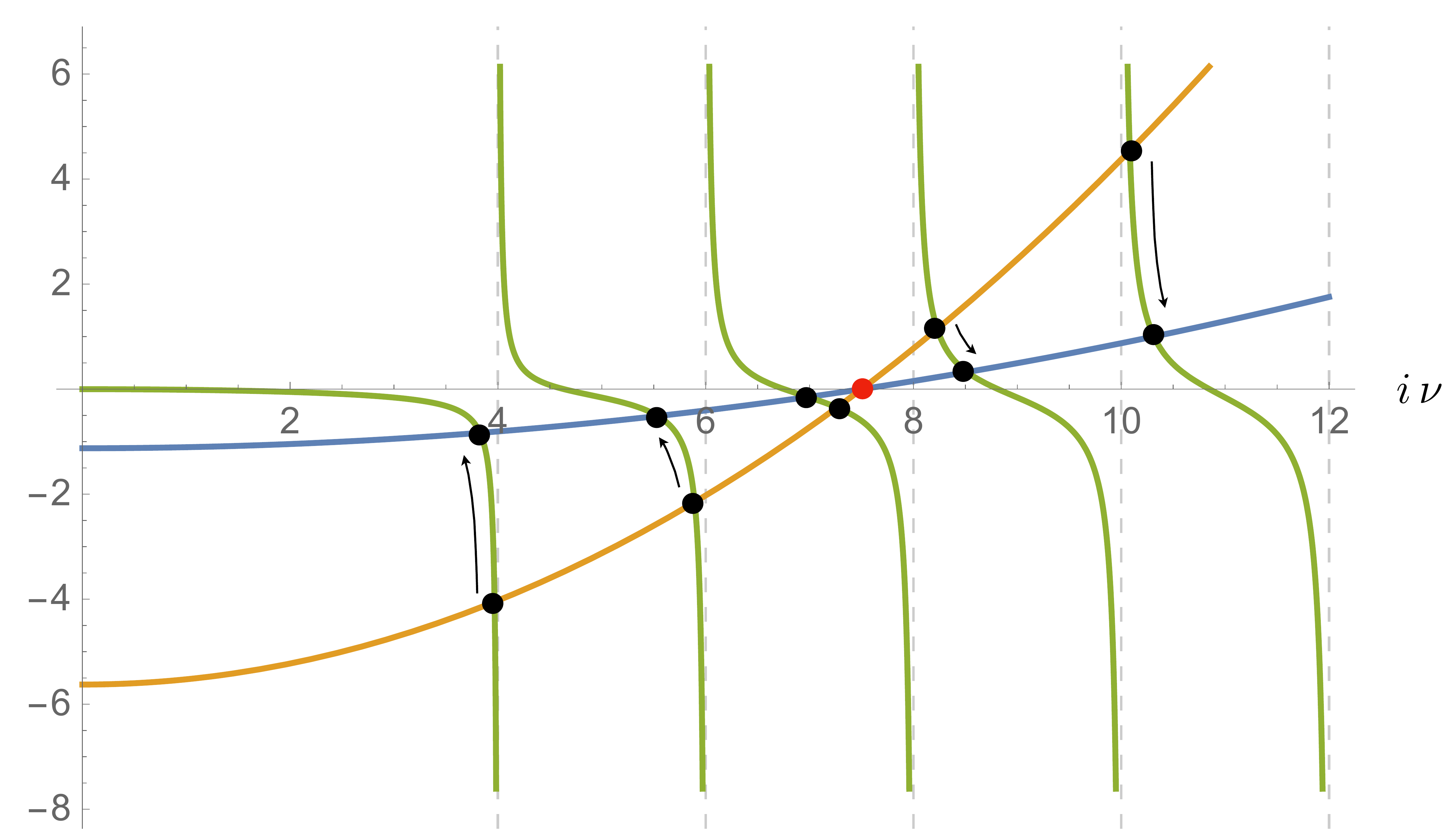}
\caption{
Plot of $B^{(1)}(\nu)$ (green line) on the negative imaginary $\nu$ axis, together with $-\frac{1}{\alpha}(\nu^2 + (\Delta_A - \tfrac{d}{2})^2)$ for $\alpha = 10$ (orange curve) and $\alpha = 50$ (blue curve) in units of the AdS radius, and $\Delta_A = 8.5$. We are in $d=2$, i.e. AdS$_3$. The intersections are highlighted with black dots, the corresponding values of $\nu$ give the scaling dimensions of spin 1 operators via $\Delta = \frac{d}{2}+i\nu$. The arrows denote the direction of increasing coupling constant $\alpha$. The dashed vertical lines correspond to the spin 1 double-trace operators in the limit $\alpha\to 0$ with $\Delta_A$ fixed. The red dot correspond to the dimension $\Delta_A$ of the non-conserved spin one operator associated to the massive vector classically. There is no operator with this scaling dimension in the interacting theory.}\label{fig:zeroesdHiggs}
\end{figure}

The spectrum of spin 1 operators is determined by the zeroes of the denominator of the photon propagator
\begin{equation}
\frac{1}{\alpha}\left(\nu^2 + (\Delta_A -\tfrac{d}{2})^2 \right)+ B^{(1)}(\nu)\vert{_{\Delta = d}} = 0~.
\end{equation}
We show the solutions $\{\nu^*_n\}$ in fig. \ref{fig:zeroesdHiggs}. They determine the scaling dimensions of the exchanged operators via $\Delta_n = \frac{d}{2}+i\nu^*_n$. Like in the Coulomb phase, the scaling dimensions approach the ones of the spin 1 double trace operators $\Delta^{(0)}_n = 2 d + 2n +1$ in the limit of small $\alpha$, and increase monotonically as we increase $\alpha$, without level crossing. 

\begin{figure}
\centering
\begin{minipage}{.45\textwidth}
\centering
\includegraphics[width=\textwidth]{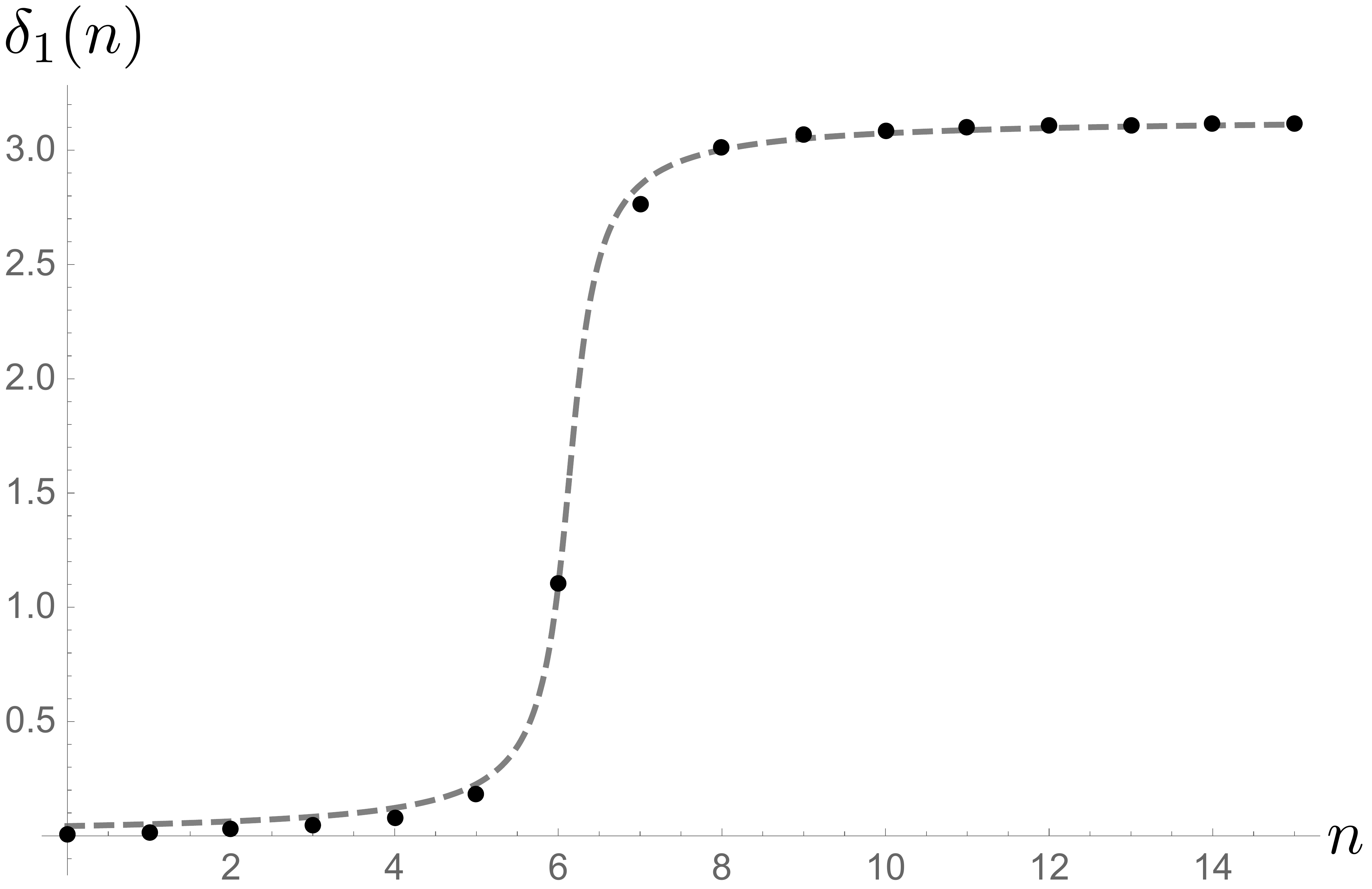}
\caption*{$\Delta_A = 16.5\,,~\alpha = 10\,,~~x =6.1\,,~y=0.3\,.$}
\end{minipage}
\hspace{1cm}\begin{minipage}{.45\textwidth}
\centering
\includegraphics[width=\textwidth]{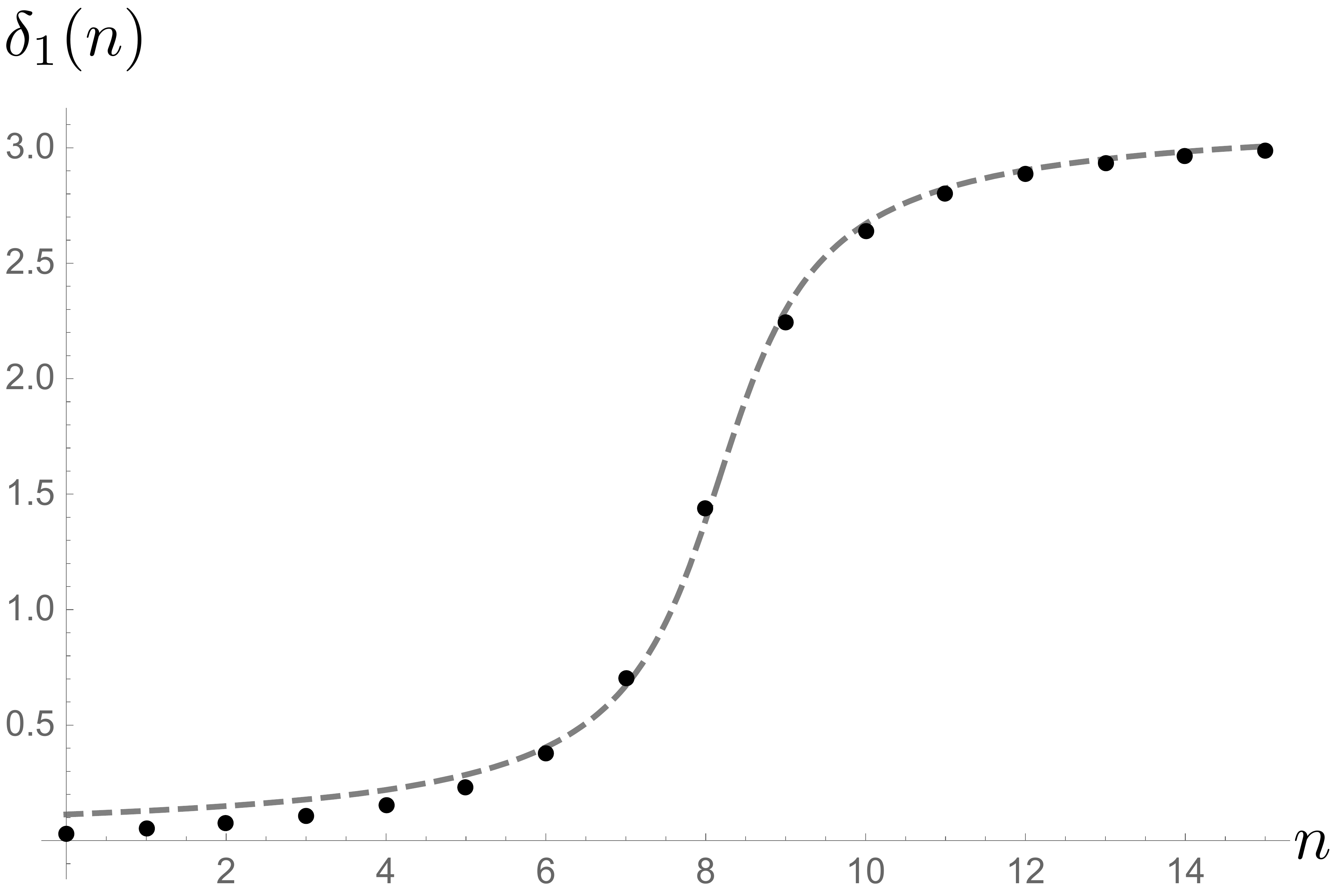}
\caption*{$\Delta_A = 20.5 \,,~\alpha = 50\,,~~x =8.2\,,~y=0.9\,.$}
\end{minipage}
\caption{The black dots are the values of \eqref{eq:phsh}, related to the anomalous dimension of the $n$-th double-trace operator, as a function of $n$ at finite coupling in the Higgs phase. The dashed line is the a fit with a Breit-Wigner phase shift, i.e. $\mathrm{Arg}\left(\frac{-1}{n- x + i y}\right)$. Both plots are for $d=2$, i.e. AdS$_3$ and the value of the parameters $\Delta_A$ and $\alpha$, as well as of the fitted parameters $x$ and $y$, are indicated under each panel. Note that as expected the resonance broadens as the coupling $\alpha$ increases.}\label{fig:phaseshift}
\end{figure}

Besides the absence of the conserved current, we see a new feature in the spectrum of the Higgs phase if we compare the anomalous dimensions $\Delta_n - \Delta^{(0)}_n$ in the two regimes $\Delta^{(0)}_n < \Delta_A$ and $> \Delta_A$. Note that contrarily to the classical expectation, in the interacting theory there is no spin 1 operator with dimension $\Delta_A$, due to the resummation of the bubble. As proposed in \cite{Paulos:2016fap} the quantity
\begin{equation}\label{eq:phsh}
\delta_{l=1} (n) = \frac{\pi}{2}\left(\Delta^{(0)}_n - \Delta_n \right)~,
\end{equation}
is related in the flat space limit to the spin 1 phase shift $\delta_{l=1}(s)$ in the scattering amplitude of the pions (the relation between $n$ and $s$ involves a certain average over a window of the discrete values of $n$ centered around the value such that $\Delta^{(0)}_n \sim \sqrt{s}$, see \cite{Paulos:2016fap} for the precise formulation of the flat space limit). The feature in $\delta_{l=1} (n)$ displayed in fig. \ref{fig:phaseshift} when $\Delta^{(0)}_n \sim \Delta_A$ is the AdS avatar of the existence of a resonance in flat space, which is characterized by a similar step behaviour of $\delta_{l=1}(s)$ around $s\sim m^2_A$.

\section{Conformal Point}\label{sec:CritAdS}

Let us now consider the limit in which $\lambda$ and $\alpha$ are sent to $+\infty$, and the mass-squared of the charged scalars is fine-tuned to have bulk conformal symmetry. Tuning the mass-squared is equivalent to fixing a particular value for the scaling dimension $\Delta$ of the boundary charged operator, so the first question to ask is what value of $\Delta$, if any, gives rise to conformal symmetry in the bulk. Note that, unlike in flat space, the correlation length is always finite in AdS, i.e. the correlation functions always decay exponentially as a function of the large geodesic distance \cite{Callan:1989em}, and there is no symmetry enhancement, so it is more subtle to detect the conformal point. Nevertheless, there are some important consequences of the bulk conformal symmetry: up to a Weyl rescaling the theory in AdS becomes equivalent to a conformal boundary condition for the CFT on a half flat-space (at least this is the case if also the boundary condition preserves conformal symmetry, more on this below). As a result there is a convergent bulk OPE expansion for correlation functions, and among the boundary operators there is a  displacement operator with protected scaling dimension $D=d+1$.

In similar setups, a criterion to detect the conformal value of a free parameter $\Delta$ from the two-point function $\langle O O \rangle$ of a bulk operator $O$ was proposed in \cite{Carmi:2018qzm}. It uses the fact that the bulk OPE expansion of the two-point function of identical operators contains the identity, and this contribution to the bulk OPE is simply a power law $\zeta^{-\Delta_O}$ of the chordal distance squared $\zeta$ that is going to zero. In a massive theory this leading power is generically accompanied by subleading integer shifted ``pseudo-descendant" powers $\zeta^{-\Delta_O+k}$, with $k\in \mathbb{N}$, but in a CFT, barring the existence of other primary operators of integer dimension $k$, these powers must be absent. It was found in examples that there exists a value of $\Delta$ setting to zero simultaneously the coefficients of all of these powers, and this determines the conformal value. This criterion can be also implemented in $\nu$ space, i.e. from the spectral representation of the two-point function: in this case one requires that the expansion at large $\nu$ matches the expansion of the spectral representation of the power-law $\zeta^{-\Delta_O}$, given by \cite{Carmi:2018qzm}
\begin{equation}
\widehat{\zeta^{-\Delta_O}}(\nu) = (4\pi)^{\frac{d+1}{2}}\frac{\Gamma(\frac{d+1}{2}-\Delta_O)}{4^{\Delta_O} \Gamma(\Delta_O)}\frac{\Gamma(-\frac{d}{2}+\Delta_O \pm i \nu)}{\Gamma(\frac12\pm i\nu)}~.
\end{equation}

Applying this criterion to the bulk two-point function of the operator $\Phi^* \Phi$ (or equivalently, the Hubbard-Stratonovich field $\sigma$) in sQED at leading order at large $N_f$, the resulting value of $\Delta$ will be identical to the one found in the $O(N)$ model, simply because at this order this two-point function is not affected by the gauge interactions. Therefore one finds that in $d=2$ the value is $\Delta=1$ \cite{Carmi:2018qzm}. More generally for any dimension $1<d<3$ we know from the study of the $O(N)$ BCFT at large $N$ in flat space \cite{McAvity:1995zd, Bray:1977tk, Ohno:1983lma} that  the conformal value is $\Delta=d-1$. We can then plug this value in the spectral representation of the large $N_f$ two-point function of the bulk gauge field, which in this limit becomes simply
\begin{align}\label{eq:propconf}
\begin{split}
 & \langle A_M(X) A_N(Y) \rangle_{\text{large $N_f$, conformal point}}  \\
& = \frac{1}{N_f}\int_{-\infty}^{+\infty} d\nu \,\frac{1}{B^{(1)}(\nu)\vert_{\Delta = d-1}} \,\Omega^{(1)}_{\nu\,MN}(X,Y) + \nabla^X_M\nabla^Y_N L(u)~,
\end{split}
\end{align}
and read-off the spectrum of spin 1 boundary operators appearing in the boundary OPE of the gauge field from the poles of $(B^{(1)}(\nu)\vert_{\Delta = d-1})^{-1}$ and also their bulk-to-boundary OPE coefficients squared from the residues. In general $d$ we cannot find these values analytically, but for any specific $d$ their numerical values can be extracted from the explicit expression \eqref{eq:Bd}, and one can check the positivity of the squared OPE coefficients.

At the integer value $d=2$, i.e. AdS$_3$, the bubble function evaluated at $\Delta=1$ simplifies to
\begin{equation}
B^{(1)}(\nu)\vert_{d=2,\Delta = 1} = \frac{\nu^3 \coth\left(\frac{\pi \nu}{2}\right)}{16(\nu^2 +1)}~,
\end{equation}
which, besides the double zero at $\nu=0$, has single zeroes at $\nu_{k,\pm} = \pm i (2k+1)$, with $k\in\mathbb{N}$ and $k\geq 1$, giving boundary operators of dimension $\Delta_k = 2k +2$. The corresponding residues are
\begin{equation}
2\pi i \mathrm{Res}_{\nu_{k,+}}\left[(B^{(1)}(\nu)\vert_{d=2,\Delta = 1})^{-1}\right] = \frac{ 256 k(k+1)}{(2k+1)^3} \geq 0~.
\end{equation}
However in this case, as we have explained in section \ref{eq:IRdiv}, the double pole at $\nu = 0$ and the associated divergence in the integral representation of the propagator imply the existence of a spin one operator of dimension 1 in the boundary spectrum, whose scalar bilinear is classically marginal and breaks conformal invariance.\footnote{It makes sense to talk about the bilinear of the operator and to sum up the scaling dimensions because we are working at large $N_f$.} Note that the operator giving rise to the boundary running coupling in the deep IR $\alpha \to \infty$ is not simply the boundary mode of the gauge field which is visible at weak coupling: recall that the boundary current is related to the bulk gauge field by $A_\mu \underset{z\to 0}{\sim} e^2 j_\mu \log z$, therefore when we take $e^2\to\infty$ to reach the IR $j_\mu$ is set to 0. Instead it is an operator from the matter sector that mixes with the gauge field at strong coupling, and that is why its properties are controlled by the bubble function.

Because of this, we cannot use Weyl rescaling to flat space and we fail to construct a conformal boundary condition for the IR fixed point of sQED with a Dirichlet boundary condition for the gauge field in AdS$_3$.  Note that this does not exclude the possibility that in flat space we might be able to define a conformal boundary condition for the IR fixed point of 3d gauge theories by starting the RG flow with a Dirichlet condition for the gauge field in the UV, because AdS and flat space with a boundary are not equivalent along the RG flow. It would be interesting to explore this question purely in flat space, for instance using the large $N_f$ expansion as a computational tool.

\section{Outlook}
\label{sec:out}

We conclude by mentioning some possible future directions:

\begin{itemize}
\item{The study of strongly coupled abelian gauge theories in AdS can be extended by considering coupling to fermionic matter and/or including also a Chern-Simons term for the gauge field in AdS$_3$. The fermionic theory also has an interesting phase diagram in AdS$_3$ which can be obtained by varying the parity-breaking but flavor-symmetry preserving mass term (or, alternatively, the parity-preserving but flavor-symmetry preserving one). At large $N_f$, by computing the fermionic current-current bubble diagram one could readily study the spectrum of operators appearing in the four-point function of the fermionic boundary operators in these phases. Note that in the massive parity-breaking phases a Chern-Simons term for the gauge field would be generated, curing the IR divergence in the photon propagator with Dirichlet condition;}
\item{Considering purely a Chern-Simons kinetic term for the gauge field, it would be interesting to study the boundary correlation functions for Chern-Simons matter theory, and to try to elucidate the unusual properties under crossing symmetry of the scattering of anyons \cite{Jain:2014nza, Gabai:2022snc, Mehta:2022lgq} from the point of view of the boundary conformal correlators;}
\item{Scattering amplitudes of charged particles in abelian gauge theories in flat space have IR divergences in $D\leq 4$. A direction for the future is to understand the AdS counterpart of the inclusive observables that give finite results, using the behavior in the flat-space limit as a diagnostic of the IR properties, see \cite{Duary:2022pyv, Duary:2022afn} for work in this direction. In particular one could compute the 1 loop diagrams in AdS that correspond to the IR divergent amplitude in flat space, and study their behavior in the flat-space limit to look for an appropriate prescription that gives a finite result. The IR divergences can also be studied at large $N_f$ and finite coupling, by computing at next-to-leading order in the $1/N_f$ expansion;}
\item{It would be interesting to try to apply bootstrap techniques to the boundary correlators of gauge theories in AdS. An important problem in this direction is to understand what are the minimal set of assumptions that allow to single out a particular gauge theory. A nice feature of the Dirichlet boundary condition is that the gauge group becomes a global symmetry at the boundary and therefore is visible in the conformal bootstrap. A natural target in this case is the four-point function of the non-abelian currents, see \cite{He:2023ewx} for recent numerical progress on this problem. Even for a fixed gauge group and matter content, in this setup one always finds not just a single conformal theory but rather a continuous family of them parametrized by the dimensionless combination of the gauge coupling and the AdS radius. Therefore important inputs for the bootstrap problem can come from the regime of weak coupling where the data of the conformal theory can be reliably computed in perturbation theory.}
\item It would be interesting to obtain explicit position-space expressions for loop diagrams with gauge fields. For this, one can use similar techniques to those obtained in \cite{Carmi:2019ocp,Carmi:2021dsn} for scalar and fermionic diagrams. Presumably there will be various relations between the spinning diagrams and the scalar diagrams.
\end{itemize}

\section*{Acknowledgements}

We thank Christian Copetti, Riccardo Ciccone, Fabiana de Cesare, Victor Gorbenko, Shota Komatsu, Ziming Ji, Manuel Loparco and Marco Serone for useful discussions and/or collaboration on related topics. We also thank the organizers of the workshops ``S-matrix bootstrap IV'' in Heraklion and ``Conformal Bootstrap, CFTs and Gravity" at LITP, Haifa where this work was presented. A. thanks ICTP for funding his doctoral scholarship. A. and L.D. are partially supported by the INFN ``Iniziativa Specifica ST\&FI''. L.D. also acknowledges support by the program ``Rita Levi Montalcini'' for young researchers.

\appendix

\section{AdS Spectral Representation for Spin 1}\label{app:SpecRep}
\subsection{Propagator and Harmonic Function}
To make the presentation self-contained, here we review the definition of the spin 1 harmonic function and the spectral representation of spin 1 propagators from \cite{Costa:2014kfa}. 

We can decompose any two-point function of spin-1 operators as
\begin{equation}\label{2ptdecmp}
F\left(X_1, X_2 ; W_1, W_2\right)=\left(W_1 \cdot W_2\right) F_0(u)+\left(W_1 \cdot X_2\right)\left(W_2 \cdot X_1\right) F_1(u)~,
\end{equation}
where $X_{1,2}$ satisfying $X_1^2 = X_2^2 = -1$, $X_{1,2}^0>0$ are embedding coordinates for points in AdS$_{d+1}$, and $W_{1,2}$ satisfying $W_{1,2}^2 = W_1\cdot X_1 = W_2 \cdot X_2 = 0$ are auxiliary vectors used to keep track of the possible structures of the  indices. For more details about embedding coordinates see \cite{Costa:2014kfa}, whose conventions we adopt here. The argument of the coefficient functions is $u = (X_1-X_2)^2/2$. In any coordinate system $x^\mu$ the two structures can be rewritten using the expression of the chordal distance in that coordinate system, and the substitutions $\left(W_1 \cdot W_2\right) \to -\frac{\partial^2 u}{\partial x_1^{\mu_1} \partial x_2^{\mu_2}}$ and $\left(W_1 \cdot X_2\right)\left(W_2 \cdot X_1\right) \to \frac{\partial u}{\partial x_1^{\mu_1} }\frac{\partial u}{\partial x_2^{\mu_2} }$.

For a massive spin 1 Proca field with $m^2 = (\Delta-1)(\Delta-d+1)$ the propagator $G_{\Delta,1}$ has the following coefficient functions
\begin{align}
\begin{split}
(G_{\Delta,1})_0\left(u\right) & =(d-\Delta) F_1(u)-\frac{1+u}{u} F_2(u), \\
(G_{\Delta,1})_1\left(u\right) & = \frac{(1+u)(d-\Delta)}{u(2+u)} F_1(u)-\frac{d+(1+u)^2}{u^2(2+u)} F_2(u)~.
\end{split}
\end{align}
where
\begin{align}
\begin{split}
 F_1(u) & =\mathcal{N}(2 u)^{-\Delta}{ }_2 F_1\left(\Delta, \frac{1-d+2 \Delta}{2}, 1-d+2 \Delta,-\frac{2}{u}\right)~, \\
 F_2(u) & =\mathcal{N}(2 u)^{-\Delta}{ }_2 F_1\left(\Delta+1, \frac{1-d+2 \Delta}{2}, 1-d+2 \Delta,-\frac{2}{u}\right)~, \\
& \mathcal{N}  =\frac{\Gamma(\Delta+1)}{2 \pi^{d / 2}(d-1-\Delta)(\Delta-1) \Gamma\left(\Delta+1-\frac{d}{2}\right)}~.
\end{split}
\end{align}
This propagator solves the equation
\begin{align}
\begin{split}\label{eq:Proca}
(-\nabla_1^2 + m^2 - d)G_{\Delta,1}(X_1,X_2; W_1,W_2) & = (W_1\cdot W_2) \delta^{d+1}(X_1,X_2)~,\\
(K_1 \cdot \nabla_1) G_{\Delta,1}(X_1,X_2; W_1,W_2) & = 0 ~.
\end{split}
\end{align}
Near the AdS boundary it satisfies a Dirichlet-type of boundary condition, namely setting $X_2 = \lambda P_2 + \mathcal{O}(\lambda^{-1})$ with $P_2^2 = 0$ the propagator scales like $\lambda^{-\Delta}$ at large $\lambda$. This is the boundary condition corresponding to the existence of a boundary operator of dimension $\Delta$ in the bulk-to-boundary OPE of the vector. Taking the limit and dividing by $\lambda^{-\Delta}$ one gets the bulk-to-boundary propagator
 \begin{align}
 \begin{split}
K_{\Delta,1}(X_1, P_2 ; W_1, Z_2) & =(d-1-\Delta) \mathcal{N} \frac{(-2 X_1\cdot P_2 )(W_1 \cdot Z_2)+2(W_1 \cdot P_2)(Z_2 \cdot X_2)}{(-2X_1\cdot P_2)^{\Delta+1}}~.
\end{split}
\end{align}

The spin 1 harmonic function can be defined in terms of the Proca propagator as
\begin{equation}\label{eq:harmdef}
\Omega_{\nu,1}\left(X_1, X_2 ; W_1, W_2\right)=\frac{i \nu}{2 \pi}\left(G_{h+i \nu,1}\left(X_1, X_2 ; W_1, W_2\right)-G_{h-i \nu,1}\left(X_1, X_2 ; W_1, W_2\right)\right)~.
\end{equation}
The coefficient functions for $\Omega_{\nu,1}$ can be written in the following form that makes manifest the absence of singularities at $u=0$
\begin{align}
\begin{split}
(\Omega_{\nu,1})_0(u) & = \frac{ \nu \sinh(\pi \nu)\left(d^2+4 \nu ^2\right) \Gamma \left(\frac{d}{2}-1+ i \nu \right)\Gamma \left(\frac{d}{2}-1 - i \nu \right) }{2^{d+4} \pi^{\frac{d+3}{2} }\Gamma \left(\frac{d+3}{2}\right)} \\
&\left[(d+1) \, _2F_1\left(\tfrac{d}{2} -i\nu,\tfrac{d}{2}+i\nu;\tfrac{d+1}{2};-\tfrac{u}{2}\right) \right.  \\
&~~~~~~\left.-(u+1) \, _2F_1\left(\tfrac{d}{2} +1-i\nu,\tfrac{d}{2}+1+i\nu;\tfrac{d+3}{2};-\tfrac{u}{2}\right)\right]~,\\
(\Omega_{\nu,1})_1(u) & = \frac{ \nu \sinh(\pi \nu)\left(d^2+4 \nu ^2\right) \Gamma \left(\frac{d}{2}-1+ i \nu \right)\Gamma \left(\frac{d}{2}-1 - i \nu \right) }{2^{d+4} \pi^{\frac{d+3}{2}  }\Gamma \left(\frac{d+3}{2}\right) u(2+u)} \\
&\left[(d+1)(1+u) \, _2F_1\left(\tfrac{d}{2} -i\nu,\tfrac{d}{2}+i\nu;\tfrac{d+1}{2};-\tfrac{u}{2}\right) \right.  \\
&~~~~~~\left.-(d+(1+u)^2) \, _2F_1\left(\tfrac{d}{2} +1-i\nu,\tfrac{d}{2}+1+i\nu;\tfrac{d+3}{2};-\tfrac{u}{2}\right)\right]~.
\end{split}
\end{align}
It is an eigenvector of the spin 1 Laplacian and transverse (i.e. divergence free)
\begin{align}
\begin{split}
-\nabla_1^2\,\Omega_{\nu, 1}& =\left(\tfrac{d^2}{4}+\nu^2+1\right) \Omega_{\nu, 1}~, \\
( K_1  \cdot \nabla_1)\Omega_{\nu, 1}& =0~.
\end{split}
\end{align}
The eigenvalues are labeled by $\nu > 0$ but it is convenient to think of $\Omega_{\nu,1}$ as an even function of $\nu \in \mathbb{R}$. It admits the so-called split representation in terms of integral of bulk-to-boundary propagators
\begin{align}
\begin{split}
&\Omega_{\nu, 1}\left(X_1, X_2 ; W_1, W_2\right) \\
& \hspace{2cm}=\frac{\nu^2}{\pi (\frac{d}{2}-1)} \int_{\partial} d P \,K_{h+i \nu, 1}\left(X_1, P ; W_1, D_Z\right) \,K_{h-i \nu, 1}\left(X_2, P ; W_2, Z\right)~.
\end{split}
\end{align}
Moreover the harmonic function satisfies the following orthogonality property under convolutions
\begin{align}\label{eq:ortho}
\begin{split}
&\frac{1}{\frac{d-1}{2}}\int_{A d S_{d+1}} d^{d+1} X_3 \,\Omega_{\nu^{\prime},1}  \left(X_1, X_3 ; W_1, K_3\right) \Omega_{\nu,1}\left(X_3, X_2 ; W_3, W_2\right) \\
& \hspace{4.5cm} =\frac{\delta\left(\nu-\nu^{\prime}\right)+\delta\left(\nu+\nu^{\prime}\right)}{2}\, \Omega_{\nu,1}\left(X_1, X_2 ; W_1, W_2\right)~,
\end{split}
\end{align}
and completeness relation 
\begin{align}\label{eq:comprel}
\begin{split}
\int_{-\infty}^{+\infty} d\nu\,\Omega_{\nu,1}(X_1,X_2;W_1,W_2) & = (W_1\cdot W_2) \delta^{d+1}(X_1,X_2) \\&  \hspace{-2cm}- (W_1\cdot \nabla_1)(W_2\cdot \nabla_2)\int_{-\infty}^{+\infty} d\nu\frac{1}{\nu^2+\frac{d^2}{4}}\,\Omega_{\nu}(X_1,X_2) ~.
\end{split}
\end{align}
Here $\Omega_{\nu}$ is the scalar harmonic function, see e.g. the appendix B of \cite{Carmi:2018qzm} and references therein.

\subsection{Spin 1 spectral representation}

We can use the spin 1 and spin 0 harmonic functions to decompose the two-point function of a spin 1 operator as follows
\begin{align}
\begin{split}\label{eq:decom2}
F(X_1,X_2;W_1,W_2) & = \int_{-\infty}^{+\infty} d\nu~\widetilde{F}^{\perp}(\nu)\,\Omega_{\nu,1}(X_1,X_2;W_1,W_2) \\ & + (W_1\cdot \nabla_1)(W_2\cdot \nabla_2) \,\int_{-\infty}^{+\infty} d\nu~\widetilde{F}^{L}(\nu)\,\Omega_\nu(X,Y)~.
\end{split}
\end{align}
Therefore we can trade the function $F$ with two functions of $\nu$, the coefficient $\widetilde{F}^{\perp}(\nu)$ of the transverse structure, and the coefficient $\widetilde{F}^{L}(\nu)$ of the longitudinal structure.

As a first example, let us consider the Proca propagator $G_{\Delta,1}$ for $\Delta > \frac{d}{2}$. Then we have \cite{Costa:2014kfa}
\begin{align}
\begin{split}
\widetilde{G_{\Delta,1}}^{\perp}(\nu) & = \frac{1}{\nu^2 +(\Delta-\frac{d}{2})^2}~, \\
\widetilde{G_{\Delta,1}}^{L}(\nu) & =  \frac{1}{(\Delta-1)(\Delta-d+1)} \frac{1}{\nu^2 + \frac{d^2}{4}}~.
\end{split}
\end{align}
This can be checked by performing explicitly the integral over the $\nu$ variable: using \eqref{eq:harmdef} to rewrite $\Omega_{\nu,1}$, we can close the contour with an arc at infinity in the upper half-plane for the $G_{\frac{d}{2}-i\nu}$ term, and in the lower half plane for the $G_{\frac{d}{2}+i\nu}$ term. The contributions of the poles at $\nu=\pm i(\Delta-\frac{d}{2})$ of $\widetilde{G_{\Delta,1}}^{\perp}(\nu)$ sum up to give precisely $G_{\Delta,1}$. In addition, there are spurious contributions from the poles of the function $G_{\frac{d}{2}\mp i\nu}$ at $\nu=\pm i (\frac{d}{2}-1)$, which are canceled from the contributions of the poles at $\nu =\pm i \frac{d}{2}$ of the longitudinal term $\widetilde{G_{\Delta,1}}^{L}(\nu)$. 

Consider now the propagator $G_{d-1,1}$ for the massless vector in the $R_\xi$ gauge with Dirichlet-type boundary conditions, giving a conserved current on the boundary. In this case we have
\begin{align}
\begin{split}
\widetilde{G_{d-1,1}}^{\perp}(\nu) & = \frac{1}{\nu^2 +(\frac{d}{2}-1)^2}~, \\
\widetilde{G_{d-1,1}}^{L}(\nu) & =   \frac{\xi}{(\nu^2 + \frac{d^2}{4})^2}~.
\end{split}
\end{align}
Note that the transverse part is precisely the limit $\Delta \to d-1$ of the transverse part of the Proca propagator. The longitudinal term instead is given by $\xi$ times the square of the $\nu$-space propagator of a massless scalar field. One can explicitly check that this expression for $G_{d-1,1}$ satisfies the correct equation
\begin{align}
\begin{split}
&\left(-\nabla_1^2 - d + (1-\tfrac{1}{\xi}) \frac{1}{\tfrac{d-1}{2}} (W_1\cdot \nabla_1) (K_1\cdot \nabla_1)\right)G_{d-1,1}(X_1,X_2; W_1,W_2) \\
&\hspace{8cm} = (W_1\cdot W_2) \delta^{d+1}(X_1,X_2)~,
\end{split}
\end{align}
by using that both $\Omega_{\nu,1}$ and $\Omega_\nu$ are eigenvalues of the Laplacian, that $\Omega_{\nu,1}$ is transverse, and the completeness relation \eqref{eq:comprel}.

In order to compute the exact propagator of the gauge field in the main text, we adopt the spectral representation for the 1PI correction to the propagator, which can be understood as the two-point correlator of the conserved current in the free theory. Motivated by this, let us now consider the two-point function of a conserved current in AdS
\begin{equation}
\left\langle J\left(X_1, W_1\right) J\left(X_2 , W_2\right)\right\rangle \equiv F_J\left(X_1, X_2 ; W_1, W_2\right)~.
\end{equation} 
Being a transverse function, its spectral representation will be in terms of a single function of $\nu$ which we will just call $\widetilde{F_J}(\nu)$, namely
\begin{equation}\label{eq:specrep1}
F_J(X_1,X_2;W_1,W_2)  = \int_{-\infty}^{+\infty} d\nu~\widetilde{F_J}(\nu)\,\Omega_{\nu,1}(X_1,X_2;W_1,W_2)~.
\end{equation} 
Thanks to \eqref{eq:ortho}, this integral transform maps convolutions to products
\begin{equation}
\widetilde{F_J\star F_{J'}}(\nu)   = \widetilde{F_J}(\nu)\widetilde{F_{J'}}(\nu)~,
\end{equation}
where
\begin{align}
\begin{split}
&(F_J\star F_{J'})(X_1,X_2;W_1,W_2)  \\ &\hspace{1cm}\equiv \frac{1}{\frac{d-1}{2}}\int_{AdS_{d+1}} d^{d+1} X_3\,F_J(X_1,X_3;W_1,K_3)F_{J'}(X_3,X_2;W_3,W_2)~.
\end{split}
\end{align} 

\subsubsection{Inversion Formula}
Let us show how to obtain the function $\widetilde{F_J}(\nu)$ from $F_J$, ``inverting'' the defining relation \eqref{eq:specrep1}.  Applying a convolution with $\Omega_{\nu,1}$ to both sides of the representation \eqref{eq:specrep1} and using the relation \eqref{eq:ortho} we get
\begin{align}
\begin{split}\label{eq:convFJ}
\frac{1}{\frac{d-1}{2}}& \int_{AdS_{d+1}} d^{d+1} X_3\, \Omega_{\nu,1}(X_1,X_3; W_1, K_3)F_J(X_3,X_2;W_3,W_2) \\ &\hspace{7cm}= \widetilde{F_J}(\nu)\,\Omega_{\nu,1}(X_1,X_2;W_1,W_2)~.
\end{split}
\end{align}
Next we compute the contraction of the two indices of  $\Omega_{\nu,1}$ and take the limit of coincident points, obtaining
\begin{align}
\begin{split}
&\frac{1}{\frac{d-1}{2}} \,\Omega_{\nu,1}\left(X_1, X_1 ; K_1, W_1\right)=C_\nu~,\\
C_\nu& \equiv\frac{d \nu\left(d^2+4 \nu^2\right) \sinh (\pi \nu) \Gamma\left(\frac{d}{2} -1 + i \nu\right)\Gamma\left(\frac{d}{2} -1 - i \nu\right)}{(4 \pi)^{\frac{d+3}{2}} \Gamma\left(\frac{d+1}{2}\right)}~.
\end{split}
\end{align}
Applying this operator to both sides of \eqref{eq:convFJ} we obtain 
\begin{align}
\begin{split}\label{eq:convFJ}
& \hspace{1cm}\widetilde{F_J}(\nu) \\  & \hspace{-0.5cm}= \frac{1}{C_\nu} \,\frac{1}{\left(\frac{d-1}{2}\right)^2}  \int_{AdS_{d+1}} d^{d+1} X_3\, \Omega_{\nu,1}(X_1,X_2; K_1, K_2)F_J(X_3,X_1;W_2,W_1)~.
\end{split}
\end{align}
This is the desired expression of $\widetilde{F_J}(\nu)$ as an integral of the function $F_J$ in position space. If we plug in the above integral the expression of both $\Omega_{\nu,1}$ and $F_J$ in the basis of structures \eqref{2ptdecmp}, we can rewrite the above integral just in terms of scalar quantities. The result is
\begin{align}
\begin{split}\label{eq:invform}
 \widetilde{F}_J(\nu)& =\frac{\mathrm{Vol}\left(S^d\right)}{C_\nu} \int_0^{+\infty} d u \sqrt{g(u)}\left\{\left(d+(1+u)^2\right) (\Omega_{\nu,1})_0(u) (F_J)_0(u)\right. \\
& -u(1+u)(2+u)\left[(\Omega_{\nu,1})_1(u) (F_J)_0(u)+(\Omega_{\nu,1})_0(u) (F_J)_0(u)\right] \\
& \left.+u^2(2+u)^2 (\Omega_{\nu,1})_1(u) (F_J)_1(u)\right\}~,
\end{split}
\end{align}
where $\sqrt{g(u)} = \frac{1}{2}(u(2+u))^{\frac{d-1}{2}}$ and $\mathrm{Vol}\left(S^d\right) = \frac{2 \pi^{\frac{d+1}{2}}}{\Gamma(\frac{d+1}{2})}$.

\subsection{Flat Space Limit}
We now show how the spectral representation discussed above reduces to the Fourier transform in the flat space limit. In order to do this we restore the dependence on the radius $L$ in the parametrization of the two-point function \eqref{2ptdecmp} by replacing $u$ with the dimensionless combination $L^{-2} u$, and allowing for an overall power of $L$ fixed by dimensional analysis. Then in the limit $L\to \infty$ the decomposition \eqref{2ptdecmp} becomes
\begin{equation}
\begin{gathered}
L^{-2 \alpha}\left(-\frac{\partial u}{\partial x^\mu \partial y^\nu} F_0\left(L^{-2} u\right)+L^{-2} \frac{\partial u}{\partial x^\mu} \frac{\partial u}{\partial y^\nu} F_1\left(L^{-2} u\right)\right) \\
\underset{L \rightarrow \infty}{\longrightarrow} \delta_{\mu \nu} f_0(u)-(x-y)_\mu(x-y)_\nu f_1(u)~,
\end{gathered}
\end{equation}
Here, $2\alpha$ can be defined as the mass dimension of the function $f_
{0}(u)$ in flat space. After taking the limit we can identify $u$ with $(x-y)^2/2$. Equivalently we can write the limit separately for the two component functions as
\begin{equation}
\begin{aligned}\label{eq:limcomp}
& F_0\left(L^{-2} u\right) \underset{L \rightarrow \infty}{=} L^{2 \alpha}\left(f_0(u)+\mathcal{O}\left(L^{-2}\right)\right)~, \\
& F_1\left(L^{-2} u\right) \underset{L \rightarrow \infty}{=} L^{2 \alpha} L^2\left(f_1(u)+\mathcal{O}\left(L^{-2}\right)\right)~.
\end{aligned}
\end{equation}
In flat space the constraint of transversality is most easily imposed in momentum space and reads
\begin{equation}
\delta_{\mu \nu} f_0(u)-(x-y)_\mu(x-y)_\nu f_1(u)=\int \frac{d^{d+1} p}{(2 \pi)^{d+1}}\left(-p^2 \delta_{\mu \nu}+p_\mu p_\nu\right) \widetilde{h}\left(p^2\right) e^{-i p(x-y)}~,
\end{equation}
which in turn implies 
\begin{equation}\label{eq:hdef}
\widetilde{h}\left(p^2\right)=-\frac{1}{p^2}\left(\widetilde{f}_0\left(p^2\right)+2 \widetilde{f}_1^{\prime}\left(p^2\right)\right)=4 \widetilde{f}_1^{\prime \prime}\left(p^2\right)~.
\end{equation}
The first of these two equalities can be seen simply as the definition of $\widetilde{h}\left(p^2\right)$ in terms of the Fourier transforms $\widetilde{f}_{0,1}\left(p^2\right)$ of $f_{0,1}(u)$, while the second expresses the constraint $f_0$ and $f_1$ need to satisfy in order to ensure transversality. Note that the primes denote derivatives w.r.t. the argument, namely $p^2$. We can then write $\widetilde{f}_{0,1}\left(p^2\right)$ as a radial Fourier transform
\begin{equation}
\tilde{f}_{0,1}\left(p^2\right)=\left(p^2\right)^{-\frac{d-1}{4}}(2 \pi)^{\frac{d+1}{2}} \int_0^{+\infty} d r \,r^{\frac{d+1}{2}} J_{\frac{d-1}{2}}\left(p r\right) f_{0,1}\left(u\right)
\end{equation}
where $p\equiv \sqrt{p^2}$, $u=r^2/2$ and $J_\nu$ is the Bessel function. Plugging in \eqref{eq:hdef} we obtain an expression for $\widetilde{h}\left(p^2\right)$ as an integral of $f_0$ and $f_1$
\begin{equation}\label{fourierflat}
\begin{aligned}
\widetilde{h}\left(p^2\right)=-\left(p^2\right)^{-\frac{d+3}{4}}(2 \pi)^{\frac{d+1}{2}} \int_0^{+\infty} d r \,r^{\frac{d+1}{2}} & \left[J_{\frac{d-1}{2}}\left(p r\right) f_0\left(u\right) -\frac{r}{p} J_{\frac{d+1}{2}}\left(p r\right) f_1\left(u\right)\right]~.
\end{aligned}
\end{equation}

Plugging in the inversion formula \eqref{eq:invform} the following identities for the asymptotic behaviors of $(\Omega_{\nu,1})_0$ and $(\Omega_{\nu,1})_1$\footnote{These identities can be derived starting from the integral expression of the hypergeometric functions entering $(\Omega_{\nu,1})_0$ and $(\Omega_{\nu,1})_1$, along the same lines of the calculation needed for the flat space limit of the scalar spectral representation in \cite{Carmi:2018qzm}.}
\begin{align}
\begin{split}
L^{-d}(\Omega_{\nu,1})_0(L^{-2} u)\vert_{\nu = p L} & \underset{ L \rightarrow \infty}{\rightarrow}\frac{(r p)^{\frac{d+1}{2}}}{2^{\frac{d+3}{2}}\pi^{\frac{d+1}{2}}r^d}  \left(J_{\frac{d-1}{2}}(r p)-\frac{1}{r p} J_{\frac{d+1}{2}}(r p)\right) ~, \\
L^{-d-2}(\Omega_{\nu,1})_1(L^{-2} u)\vert_{\nu = p L} & \underset{ L \rightarrow \infty}{\rightarrow} \frac{(r p)^{\frac{d+1}{2}}}{2^{\frac{d+3}{2}}\pi^{\frac{d+1}{2}}r^{d+2}} \left(J_{\frac{d-1}{2}}(r p)-\frac{d+1}{r p} J_{\frac{d+1}{2}}(r p)\right) ~,
\end{split}
\end{align}
together with the limits \eqref{eq:limcomp}, and comparing with \eqref{fourierflat}, we get
\begin{equation}
L^{-2 \alpha+d+1}\widetilde{F}_J(\nu) \vert_{\nu = p L}  \underset{ L \rightarrow \infty}{\rightarrow}  p^2 \widetilde{h}\left(p^2\right)~.
\end{equation}
This is the desired relation between the spectral representation and the Fourier transform in the flat space limit.

\bibliographystyle{JHEP}
\bibliography{QEDAdSN}


\end{document}